\documentclass[12pt]{article}
\pdfoutput=1
\usepackage{putex}
\usepackage{graphicx}
\usepackage{epstopdf}
\usepackage{cite}
\usepackage{tensor}
\usepackage{breqn}
\usepackage{slashed}
\usepackage{hyperref}
\usepackage{float}
\usepackage[bottom]{footmisc}
\usepackage[utf8]{inputenc}
\usepackage{amsmath}
\usepackage[dvipsnames,svgnames,table]{xcolor}
\usepackage{subcaption}

\usepackage{tikz}
\usetikzlibrary{positioning,calc}
\usetikzlibrary{decorations.pathmorphing}
\usetikzlibrary{decorations.markings, decorations.pathmorphing}
\tikzset{graviton/.style={
	decorate,
	decoration={snake, amplitude=1, segment length=5},
}}
\usetikzlibrary{arrows.meta}

\definecolor{rnblue}{RGB}{27,99,137}
\definecolor{shellblue}{RGB}{82,145,214}

\numberwithin{equation}{section}
\hypersetup{
	pdfstartview={FitH},    
	pdftitle={Enhanced Correlations in Hawking Radiation from Near-Extremal Collapse  },    
	pdfauthor={Bintanja, Kim, Kraus},     
	colorlinks=true,       
	linkcolor=blue,          
	citecolor=blue!59!black! ,        
	filecolor=magenta,      
	urlcolor=blue!75!black,           
	linktoc=all
}

\makeatletter
\g@addto@macro\bfseries{\boldmath}
\makeatother

\numberwithin{equation}{section}

\newcommand{\eq}[2]{\begin{align}\label{#1}#2\end{align}}

\def\Hc{{\cal H}}
\newcommand {\be} {\begin {equation}}
\newcommand {\ee} {\end {equation}}

\newcommand{\p}{\partial}

\newcommand\rt{{\rightarrow}}
\def\eps{\epsilon}

\newcommand{\rf}[1]{(\ref{#1})}
\newcommand{\rff}[1]{\ref{#1}}

\newcommand{\phit}{\tilde{\phi}}

\newcommand{\Oc}{{\cal O}}

\newcommand{\phid}{\dot{\phi}}

\newcommand{\mathscr}{\mathcal}

\newcommand{\Ic}{\mathcal{I}}

\newcommand{\ft}{\tilde{f}}
\newcommand{\veps}{\varepsilon}

\newcommand{\Rc}{{\cal R}} 
\newcommand{\Mc}{{\cal M}} 
\newcommand{\RN}{Reissner-Nordstr\"om~}
\newcommand{\ttil}{\tilde{t}} 
\newcommand{\pih}{\hat{\pi}} 
\newcommand{\Tt}{\tilde{T}} 
\newcommand{\Vt}{\tilde{V}} 
\newcommand{\Wt}{\tilde{W}} 
\newcommand{\Qc}{{\cal Q}}
\newcommand{\that}{\hat{t}}
\newcommand{\omt}{\tilde{\omega}}

\begin{document}

	\institution{UCLA}{ \quad\quad\quad\quad\quad\quad\quad\ ~ \, $^{1}$Mani L. Bhaumik Institute for Theoretical Physics
		\cr Department of Physics \& Astronomy,\,University of California,\,Los Angeles,\,CA\,90095,\,USA}

    \institution{CERN}{ \quad\quad\quad\ ~ \, $^{2}$CERN, Theory Division,
Geneva 23, CH-1211, Switzerland}

	\title{\LARGE Enhanced Correlations in Hawking Radiation \\[-1cm] from Near-Extremal Collapse   \\
    \normalsize
	}
	
	\authors{Suzanne Bintanja$^{1,2}$,  Seolhwa Kim$^{1}$,  Per Kraus$^{1}$}
	
	\abstract{We consider the formation of a near-extremal \RN black hole by collapse, and show how to compute correlations in the outgoing Hawking radiation due to enhanced gravitational backreaction effects in the near-horizon region. This is done by reducing to the s-wave and employing the Hamiltonian formulation of Einstein-Maxwell theory coupled to a scalar field. Solving the constraints yields an action for the scalar field that incorporates gravitational backreaction effects at the quantum level, governed by an effective coupling $g = G/(  \pi r_0^3T_H)$ that grows at low temperature, as in recent Schwarzian-based analyses.   This action produces corrections to the free field Hawking state which are imprinted on correlation functions of the Hawking radiation measured at null infinity. As part of our analysis, we show that this action evaluated in the AdS$_2$ region is equivalent, at the level of all tree-level boundary correlators, to the standard  JT/Schwarzian description coupled to dressed bilocal operators. We also reproduce some one-loop results. In our approach, metric fluctuations are included quantum mechanically through the reduced scalar action, rather than through a semiclassical expectation value, and our computation of the radiation manifestly reduces to Hawking's original treatment when metric fluctuations are neglected.   }
	
	\date{}

	\maketitle
	\setcounter{tocdepth}{2}
	\begingroup
	\hypersetup{linkcolor=black}
	\tableofcontents
	\endgroup

\section{Introduction}

Figure \rff{penrose} depicts a charged spherical null shell undergoing gravitational collapse to form a \RN black hole.\footnote{As shown, the solution inside the horizon is unphysical: the stress tensor of the shell violates the null energy condition as it approaches the singularity, and a physical shell instead undergoes a bounce. See \cite{Ori_1991,Horowitz:2022ptw,Gralla:2025gzl}  for discussion (we thank Sam Gralla for pointing this out and for related discussions). We set this issue aside, since our analysis only uses the exterior region.} Hawking \cite{Hawking:1975vcx} famously showed that a free quantum field on this background, prepared in a broad class of states on $\Ic^-$, including the vacuum, appears thermal at a temperature $T_H$ when probed by late-time observables on $\Ic^+$. Much effort since then has gone into thinking about how this picture changes in a full theory of quantum gravity.        

\begin{figure}[H]
\centering
\scalebox{0.80}{%
\begin{tikzpicture}[
    x=0.72cm,
    y=0.72cm,
    line cap=butt,
    line join=miter,
    every node/.style={inner sep=1pt}
]
    \coordinate (bottomleft)  at (0,0);
    \coordinate (shellleft)   at (0,11);
    \coordinate (horizonleft) at (0,5.5);
    \coordinate (uppercorner) at (5,10.5);
    \coordinate (rightcorner) at (7.75,7.75);
    \coordinate (upperleft)   at (2.6,12.9);
    \coordinate (shellright)  at (5.5,5.5);

    \draw[black, line width=1.25pt]
        (bottomleft) -- (rightcorner) -- (upperleft);

    \draw[blue, line width=1.25pt]
        (horizonleft) -- (uppercorner);

    \draw[black, dashed, line width=1.25pt]
        (bottomleft) -- (shellleft);

    \draw[black, line width=3.2pt]
        (shellleft) -- (0,12.9);

    \draw[black, line width=3.2pt]
        (uppercorner) -- (5,12.9);

    \draw[
        red,
        line width=2.35pt,
        postaction={
            decorate,
            decoration={
                markings,
                mark=at position 0.78 with {
                    \arrow{Stealth[length=4.2mm,width=3.2mm]}
                }
            }
        }
    ]
        (shellright) -- (shellleft);

    \draw[
        -{Stealth[length=3.2mm,width=2.3mm]},
        line width=1.15pt
    ]
        (8.05,8.05) -- (5.85,10.25);

    \node[
        font=\large,
        anchor=south east
    ]
        at (5.88,10.20) {$u$};

    \draw[
        -{Stealth[length=3.2mm,width=2.3mm]},
        line width=1.15pt
    ]
        (0.72,0.10) -- (7.55,6.93);

    \node[
        font=\large,
        anchor=south west
    ]
        at (7.60,6.9) {$v$};

    \node[
        red,
        font=\sffamily\normalsize,
        anchor=west,
        rotate=-45
    ]
        at (1.4,10.3) {shell};

    \node[
        black,
        font=\large,
        anchor=east
    ]
        at (5.25,5.55) {$v_s$};

    \node[
        font=\sffamily\footnotesize,
        anchor=west
    ]
        at (3.05,8.15) {Reissner--Nordstr\"om};

    \node[
        font=\sffamily\footnotesize,
        anchor=west
    ]
        at (7.05,9.7) {$\Ic^+$};

    \node[
        font=\sffamily\footnotesize,
        anchor=west
    ]
        at (5.0,3.7) {$\Ic^-$};

    \node[
        blue,
        font=\sffamily\small,
        anchor=west,
        rotate=45
    ]
        at (0.5,6.55) {horizon};

    \node[
        blue,
        font=\small,
        anchor=west,
        rotate=45
    ]
        at (1.15,5.95) {$U=U_h$};

    \node[
        font=\sffamily\footnotesize,
        anchor=west
    ]
        at (1.35,4.85) {flat space};

    \node[
        font=\normalsize,
        anchor=east
    ]
        at (-0.30,3.10) {$r=0$};
       
\end{tikzpicture}%
}
\caption{Collapsing null shell forming a \RN black hole.}
\label{penrose}
\end{figure}
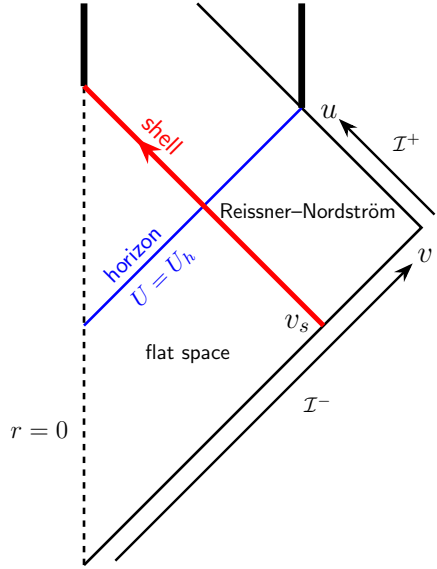

As long as the black hole is large, so that all curvature scales on and outside the horizon are small, there would appear to be no obstacle to using perturbation theory to compute corrections to the outgoing radiation, due to gravitational or other weakly coupled interactions; it is a potentially messy but fundamentally straightforward problem in time-dependent perturbation theory. In general, such corrections are not expected to be particularly interesting. An exception occurs in the case that the black hole is formed near extremality, where $T_H$ is small in a sense made precise below.  In the near-extremal regime, the fixed-background approximation in the near-horizon region begins to fail: large-scale quantum fluctuations become important even though all curvature scales remain far below the Planck scale.   This surprising fact was originally inferred from thermodynamic considerations in \cite{Preskill:1991tb}. More recently, this phenomenon has been understood in terms of the physics of the near-horizon AdS$_2$ region that develops in the extremal limit \cite{Almheiri:2014cka,Jensen:2016pah,Maldacena:2016upp,Engelsoy:2016xyb,Stanford:2017thb}, and the subsequent study of JT gravity \cite{Jackiw:1984je,Teitelboim:1983ux} as a model of AdS$_2$ has led to impressive progress in various aspects of quantum gravity.  See, e.g., \cite{Mertens:2022irh} for a review.

It is therefore interesting to compute Hawking radiation in the near-extremal regime, since this requires understanding genuine quantum gravity effects in a setting largely disentangled from the usual ultraviolet issues.  A careful analysis reveals that near-extremal Hawking radiation develops parametrically enhanced correlations governed by the dimensionless parameter $g =  {G\over \pi r_0^3 T_H}$   ($r_0$ being the horizon radius at extremality), so the free field approximation fails in a controlled and computable way (until $g$ becomes large).

The near-horizon region is indicated in Figure \rff{fig:collapse-penrose-rcurve}, and the problem is to propagate the initial quantum state of the scalar field on $\Ic^-$  through the fluctuating near-horizon region and out to $\Ic^+$, (tracing over the part of the state that falls through the horizon.)   One approach to studying related processes involving near-extremal black holes, employed in a number of works \cite{Brown:2024ajk,Emparan:2025sao,Lin:2025wof,Emparan:2025qqf,Biggs:2025nzs,Betzios:2025sct},  is to replace the near-horizon region by a Schwarzian quantum mechanics system, which then couples to fields in the exterior asymptotically flat region. This is in the same spirit as works that used D-branes to model the near-horizon region of various types of black holes in string theory, and that led to the discovery of the AdS/CFT correspondence \cite{Maldacena:1997re}.\footnote{Though the analogy is not precise, since the Schwarzian theory is not a dual description, but rather an effective description of the near-horizon region,  expressed in terms of the original gravitational variables. } Applied to the problem at hand, this approach has some advantages and disadvantages that will be discussed later. Here we take a different route, proceeding essentially as in Hawking’s original derivation,  by employing standard quantum field theory methods to evolve the quantum state from the far past to the far future,  except that we allow the spacetime geometry to fluctuate.\footnote{We distinguish this from the large body of work that uses a semi-classical approach to study backreaction, e.g., \cite{Callan:1992rs}. This approach treats the matter quantum mechanically but the metric classically, as can be justified in the large $N$ limit, $N$ denoting the number of species of matter fields.  By contrast, here we are treating the matter and metric on the same footing.} To the extent that their domains of applicability overlap, our results will turn out to mesh nicely with those of the Schwarzian approach.  Indeed, one output of our analysis will be to derive aspects of the Schwarzian approach from a different starting point, one that avoids the use of Euclidean path integrals and certain other subtleties to be discussed.   

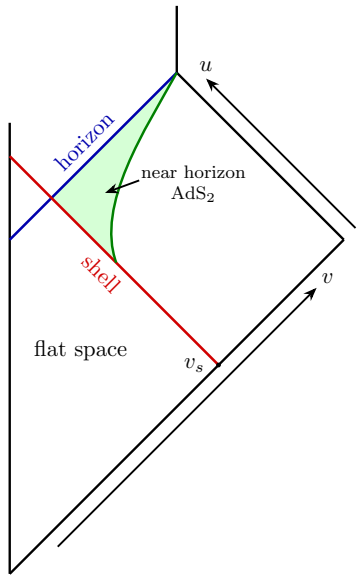
\begin{figure}[h]
\centering
\scalebox{0.85}{%
\begin{tikzpicture}[
    x=0.65cm,
    y=0.65cm,
    line cap=butt,
    line join=round,
    boundary/.style={black,line width=1.05pt},
    horizonline/.style={blue!68!black,line width=1.15pt},
    shellline/.style={red!82!black,line width=1.15pt},
    rcurve/.style={green!50!black,line width=1.10pt},
    axisarrow/.style={black,line width=0.85pt,-{Stealth[length=2.1mm,width=1.6mm]}},
    labelbox/.style={fill=white,fill opacity=0.92,text opacity=1,inner sep=1.3pt}
]

\useasboundingbox (-0.70,-0.45) rectangle (8.95,14.15);

\coordinate (Lbot) at (0,0);
\coordinate (Ltop) at (0,10.8);

\coordinate (Top) at (4,12);
\coordinate (TopUp) at (4,13.6);
\coordinate (R) at (8,8);

\coordinate (BlueStart) at (0,8);
\coordinate (ShellStart) at (0,10);
\coordinate (ShellEnd) at (5,5);

\coordinate (HSint) at (1,9);

\coordinate (RconstEnd) at (2.55,7.45);

\fill[green!16]
  (HSint) -- (Top)
  .. controls (3.25,10.65) and (2.05,8.85) ..
  (RconstEnd) -- cycle;

\draw[boundary] (Lbot) -- (Ltop);
\draw[boundary] (Lbot) -- (R);
\draw[boundary] (Top) -- (R);
\draw[boundary] (Top) -- (TopUp);

\draw[horizonline] (BlueStart) -- (Top);
\draw[shellline] (ShellStart) -- (ShellEnd);

\draw[rcurve]
  (Top)
  .. controls (3.25,10.65) and (2.05,8.85) ..
  (RconstEnd);

\path (BlueStart) -- (Top)
  node[
    pos=0.52,
    sloped,
    above=2.8pt,
    blue!68!black,
    font=\footnotesize,
    labelbox
  ] {horizon};

\path (ShellStart) -- (ShellEnd)
  node[
    pos=0.50,
    sloped,
    below=2.8pt,
    red!82!black,
    font=\footnotesize,
    labelbox
  ] {shell};

\node[
    font=\scriptsize,
    align=center,
    fill=white,
    fill opacity=0.92,
    text opacity=1,
    inner sep=1.2pt
] at (4.40,9.35) {near horizon\\AdS$_2$};

\draw[axisarrow] (3.10,9.42) -- (2.28,9.12);

\node[
    font=\footnotesize,
    fill=white,
    fill opacity=0.92,
    text opacity=1,
    inner sep=1.3pt
] at (1.70,5.35) {flat space};

\fill[black] (ShellEnd) circle[radius=1.05pt];
\node[font=\footnotesize, left=2pt] at (ShellEnd) {$v_s$};

\draw[axisarrow] (8.28,8.28) -- (4.70,11.86);
\node[font=\footnotesize,above=2pt] at (4.70,11.70) {$u$};

\draw[axisarrow] (1.15,0.65) -- (7.35,6.85);

\node[
  font=\footnotesize,
  anchor=south west,
  xshift=0pt,
  yshift=-1pt
] at (7.25,6.8) {$v$};

\end{tikzpicture}%
}
\caption{Near extremality, the near-horizon region, shaded in green, undergoes large-scale quantum fluctuations controlled by the dimensionless coupling $g = {G\over \pi r_0^3 T_H}$.}
\label{fig:collapse-penrose-rcurve}
\end{figure}

Attacking this problem requires making a judicious set of approximations.  Here we are guided by the goal of isolating the physics that gives rise to large effects near extremality. The first approximation is that we reduce the four-dimensional Einstein-Maxwell-scalar theory to the s-wave, the justification being that the large quantum fluctuations in the near-horizon region lie in the s-wave sector.  Next, we work in perturbation theory around the spacetime background depicted in Fig.  \rff{penrose}, and we only pay attention to quantum backreaction effects outside the shell, since no interesting effects are expected to arise in the flat space interior region. The shell is taken to follow a fixed trajectory in the background solution, and we quantize the metric outside the shell, not the shell itself.  Again, the justification is that the effects we are after have to do with fluctuations of the near-horizon geometry outside the shell.  Under these assumptions, we proceed by applying standard and well-established rules of quantum field theory, namely time-dependent perturbation theory in Lorentzian signature.   

Building on the previous work \cite{Kraus:2025efu}, since gravity in the s-wave has no propagating degrees of freedom, we can eliminate it by choosing a gauge and solving the gravitational constraints.  The result is an action that involves only the scalar field $\phi$, while still encoding all s-wave gravitational effects \cite{Berger:1972pg,Unruh:1976db}.  We emphasize that this procedure uses only the constraints and not the equations of motion, and so no approximation has been made other than a reduction to the s-wave. It is analogous to quantizing Yang-Mills theory by choosing axial gauge and solving the constraints, thereby arriving at a Hamiltonian for the physical degrees of freedom.  This is valid at the quantum mechanical level, and is equivalent to, say, using the Faddeev-Popov procedure in a covariant gauge.  The resulting action is written in phase space and is given by 
\eq{z1}{ S  = \int\! dt \int_{r_s}^\infty\! dr  \left( \pi_\phi \dot{\phi} -\Big(f(r)h(r) + r^2V\big(\phi(r)\big)\Big)  e^{-\int_{r}^\infty \! dr' {2 Gh(r') \over r'} }   \right) }
where $ f(r) = 1-{2GM\over r}+{GQ^2\over r^2}$ is the standard function appearing in the \RN metric,  $h(r) = {1\over 2} \left( {\pi_\phi^2 \over r^2} +r^2 {\phi'}^2\right)$, and we have included an arbitrary scalar potential $V(\phi)$. The lower limit of the $r$ integration is given by the shell location, in line with our comment above: inside the shell we take the scalar field to be free. Note that this action is nonlocal in space but local in time, the latter implying no basic obstacle to quantization apart from UV issues arising from collisions of $h(r)$ factors.  

Part of our gauge fixing involves identifying the coordinate $r$ with the proper radius of the $S^2$, i.e., with the dilaton in a JT description;  this gives $r$ a clear physical meaning.  The second gauge condition involves setting one of the gravitational canonical momenta to zero, which effectively makes $t$ behave like Schwarzschild time.\footnote{Schwarzschild time, of course, breaks down at the horizon, but this is no concern since all of our analysis will take place outside the horizon. }   

To gain some intuition for \rf{z1} we first note that if we omit the exponential factor then the action reduces to that of a free scalar in a fixed \RN background.  To go beyond this,  it is helpful to note that if we define the metric $ds^2 =-N^2 dt^2 + L^2 dr^2 + r^2 d\Omega^2$, where $N$ and $L$ are given by specific functionals of the phase space coordinates $(\phi,\pi_\phi)$, then the Euler-Lagrange equations derived from \rf{z1} are equivalent to the equations of motion of a scalar field with potential $V(\phi)$ propagating in this metric; see section \rff{equiv} for details.  The specific form of $L$  is obtained by solving one of the gravitational constraints, while the form of the lapse function $N$, which acts as a Lagrange multiplier in the Hamiltonian formulation, involves a choice.   Since the metric, or more precisely the function $L$, is tied to the scalar field via the constraints, it is clear that quantization of the scalar field implies a quantization of the metric as well. 

The action \rf{z1} may be used to compute gravitationally corrected correlation functions of the Hawking radiation on $\Ic^+$, where by correlation functions we mean in-in expectation values. We focus on two-point and four-point correlators computed to first order in Newton's constant $G$, with the corresponding diagrams depicted in Fig. \rff{fig:vertices-pair}.


\begin{figure}[h]
\centering

\begin{subfigure}{0.35\textwidth}
\centering
\resizebox{0.85\linewidth}{!}{%
\begin{tikzpicture}[
    x=0.65cm,
    y=0.65cm,
    line cap=butt,
    line join=round,
    boundary/.style={black,line width=1.05pt},
    horizonline/.style={blue!68!black,line width=1.15pt},
    shellline/.style={red!82!black,line width=1.15pt},
    rcurve/.style={green!50!black,line width=1.10pt},
    scalarline/.style={black,line width=0.95pt},
    graviton/.style={
        black,
        line width=0.85pt,
        decorate,
        decoration={snake,amplitude=1.15pt,segment length=4.5pt}
    },
    vertex/.style={black,fill=black,circle,inner sep=1.05pt},
    axisarrow/.style={black,line width=0.85pt,-{Stealth[length=2.1mm,width=1.6mm]}},
    labelbox/.style={fill=white,fill opacity=0.92,text opacity=1,inner sep=1.3pt}
]

\useasboundingbox (-0.70,-0.45) rectangle (8.95,14.15);

\coordinate (Lbot) at (0,0);
\coordinate (Ltop) at (0,10.8);

\coordinate (Top) at (4,12);
\coordinate (TopUp) at (4,13.6);
\coordinate (R) at (8,8);

\coordinate (BlueStart) at (0,8);
\coordinate (ShellStart) at (0,10);
\coordinate (ShellEnd) at (5,5);

\coordinate (HSint) at (1,9);

\coordinate (RconstEnd) at (2.55,7.45);

\coordinate (MeetPtA) at (3.0,8.8);
\coordinate (ScriA1)  at (4.8,11.2);
\coordinate (ScriA2)  at (5.5,10.5);

\coordinate (MeetPtB) at (4.2,7.6);
\coordinate (ScriB1)  at (5.8,10.2);
\coordinate (ScriB2)  at (6.5,9.5);

\draw[boundary] (Lbot) -- (Ltop);
\draw[boundary] (Lbot) -- (R);
\draw[boundary] (Top) -- (R);
\draw[boundary] (Top) -- (TopUp);

\draw[horizonline] (BlueStart) -- (Top);
\draw[shellline] (ShellStart) -- (ShellEnd);

\draw[scalarline] (MeetPtA) -- (ScriA1);
\draw[scalarline] (MeetPtA) -- (ScriA2);

\draw[scalarline] (MeetPtB) -- (ScriB1);
\draw[scalarline] (MeetPtB) -- (ScriB2);

\draw[graviton] (MeetPtA) -- (MeetPtB);

\node[vertex] at (MeetPtA) {};
\node[vertex] at (MeetPtB) {};

\path (BlueStart) -- (Top)
  node[
    pos=0.52,
    sloped,
    above=2.8pt,
    blue!68!black,
    font=\footnotesize,
    labelbox
  ] {horizon};

\path (ShellStart) -- (ShellEnd)
  node[
    pos=0.50,
    sloped,
    below=2.8pt,
    red!82!black,
    font=\footnotesize,
    labelbox
  ] {shell};

\node[
    font=\footnotesize,
    fill=white,
    fill opacity=0.92,
    text opacity=1,
    inner sep=1.3pt
] at (1.50,5.55) {flat space};

\fill[black] (ShellEnd) circle[radius=1.05pt];
\node[font=\footnotesize,left=2pt] at (ShellEnd) {$v_s$};

\draw[axisarrow] (8.28,8.28) -- (4.70,11.86);
\node[font=\footnotesize,above=2pt] at (4.70,11.70) {$u$};

\draw[axisarrow] (1.15,0.65) -- (7.35,6.85);

\node[
  font=\footnotesize,
  anchor=south west,
  xshift=0pt,
  yshift=-1pt
] at (7.25,6.8) {$v$};

\end{tikzpicture}%
}
\caption{Connected four-point function.}
\label{fig:vertices-left}
\end{subfigure}
\hspace{0.03\textwidth}
\begin{subfigure}{0.35\textwidth}
\centering
\resizebox{0.85\linewidth}{!}{%
\begin{tikzpicture}[
    x=0.65cm,
    y=0.65cm,
    line cap=butt,
    line join=round,
    boundary/.style={black,line width=1.05pt},
    horizonline/.style={blue!68!black,line width=1.15pt},
    shellline/.style={red!82!black,line width=1.15pt},
    rcurve/.style={green!50!black,line width=1.10pt},
    scalarline/.style={black,line width=0.95pt},
    graviton/.style={
        black,
        line width=0.85pt,
        decorate,
        decoration={snake,amplitude=1.15pt,segment length=4.5pt}
    },
    vertex/.style={black,fill=black,circle,inner sep=1.05pt},
    axisarrow/.style={black,line width=0.85pt,-{Stealth[length=2.1mm,width=1.6mm]}},
    labelbox/.style={fill=white,fill opacity=0.92,text opacity=1,inner sep=1.3pt}
]

\useasboundingbox (-0.70,-0.45) rectangle (8.95,14.15);

\coordinate (Lbot) at (0,0);
\coordinate (Ltop) at (0,10.8);

\coordinate (Top) at (4,12);
\coordinate (TopUp) at (4,13.6);
\coordinate (R) at (8,8);

\coordinate (BlueStart) at (0,8);
\coordinate (ShellStart) at (0,10);
\coordinate (ShellEnd) at (5,5);

\coordinate (HSint) at (1,9);

\coordinate (RconstEnd) at (2.55,7.45);

\coordinate (MeetPt) at (3.7,7.8);
\coordinate (ScriA)  at (5.6,10.4);
\coordinate (ScriB)  at (6.8,9.2);

\coordinate (PropA) at ($(MeetPt)!0.50!(ScriA)$);
\coordinate (PropB) at ($(MeetPt)!0.50!(ScriB)$);

\draw[boundary] (Lbot) -- (Ltop);
\draw[boundary] (Lbot) -- (R);
\draw[boundary] (Top) -- (R);
\draw[boundary] (Top) -- (TopUp);

\draw[horizonline] (BlueStart) -- (Top);
\draw[shellline] (ShellStart) -- (ShellEnd);

\draw[scalarline] (MeetPt) -- (ScriA);
\draw[scalarline] (MeetPt) -- (ScriB);

\draw[graviton] (PropA) -- (PropB);

\fill[black] (MeetPt) circle[radius=1.0pt];
\node[vertex] at (PropA) {};
\node[vertex] at (PropB) {};

\path (BlueStart) -- (Top)
  node[
    pos=0.52,
    sloped,
    above=2.8pt,
    blue!68!black,
    font=\footnotesize,
    labelbox
  ] {horizon};

\path (ShellStart) -- (ShellEnd)
  node[
    pos=0.50,
    sloped,
    below=2.8pt,
    red!82!black,
    font=\footnotesize,
    labelbox
  ] {shell};

\node[
    font=\footnotesize,
    fill=white,
    fill opacity=0.92,
    text opacity=1,
    inner sep=1.3pt
] at (1.50,5.55) {flat space};

\fill[black] (ShellEnd) circle[radius=1.05pt];
\node[font=\footnotesize,left=2pt] at (ShellEnd) {$v_s$};

\draw[axisarrow] (8.28,8.28) -- (4.70,11.86);
\node[font=\footnotesize,above=2pt] at (4.70,11.70) {$u$};

\draw[axisarrow] (1.15,0.65) -- (7.35,6.85);

\node[
  font=\footnotesize,
  anchor=south west,
  xshift=0pt,
  yshift=-1pt
] at (7.25,6.8) {$v$};

\end{tikzpicture}%
}
\caption{One-loop correction to the two-point function.}
\label{fig:vertices-right}
\end{subfigure}

\caption{Examples of correlation-function diagrams in the collapse geometry. We have indicated the interactions as being mediated by graviton exchange, although in our actual setup, in which gravity has been integrated out, these are represented by quartic scalar field vertices.}
\label{fig:vertices-pair}
\end{figure}
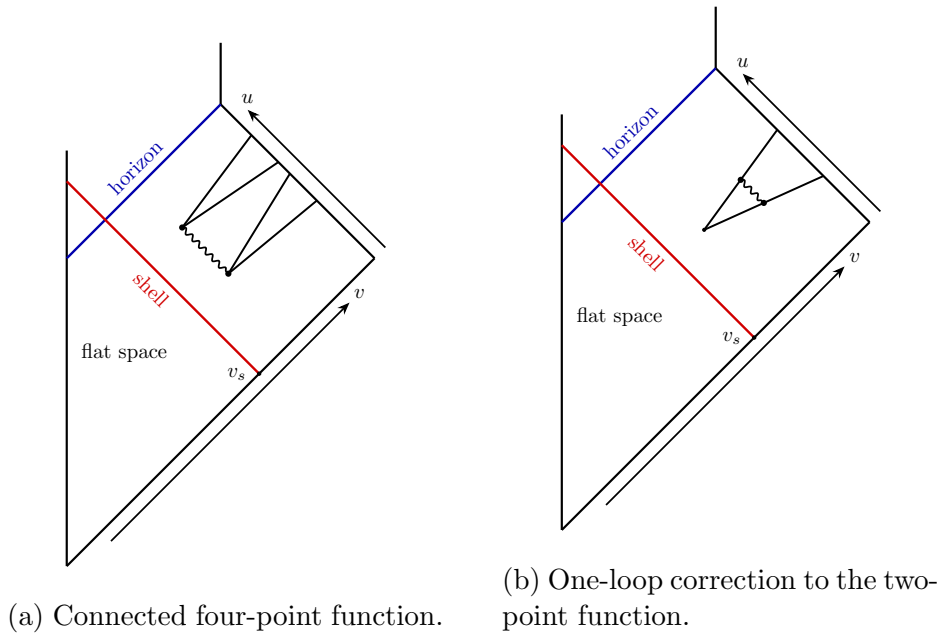


In general, the interaction vertices can be located anywhere in the spacetime, but in the near-extremal regime the dominant effects come when the vertices lie in the near-horizon AdS$_2$ region. A standard rescaling of variables reveals that the effective dimensionless coupling governing gravitational interactions in this region is given by 
\eq{z2}{ g =  {G\over  \pi r_0^3T_H}~. }
Here $r_0$ is the radius of the black hole in the extremal limit, related to the charge $Q$ as $r_0  = \sqrt{G}Q $. This is the same coupling as can be inferred from the original thermodynamic reasoning in \cite{Preskill:1991tb}, and that appears in Schwarzian-based analyses; see also \cite{Almheiri:2016fws,Nayak:2018qej,Moitra:2018jqs} for derivations based on gravitational perturbation theory.    To isolate the enhanced near-extremal interactions, we proceed by evaluating correlation functions at the outer boundary of the AdS$_2$ region and then use the free field equations to propagate them to $\Ic^+$.   This last step involves the usual greybody factors, though we mostly do not include them explicitly since they do not affect the near-horizon enhancement central to our analysis.

In principle we can evaluate these correlators at early times before the black hole has quasi-equilibrated, but here we confine our attention to obtaining explicit answers at late times,\footnote{Of course, by ``late time" we mean a time at which the black hole has reached quasi-equilibrium.  If we wait too long, then the black hole will decay so close to extremality that the effective coupling $g$  becomes large, invalidating our perturbative computation.} when the radiation is thermal in the free field approximation.  In this regime, we can relate the correlators we want to thermal AdS$_2$ correlators.  Using the quartic interaction vertex obtained by expanding the exponential in \rf{z1} to first order, we can compute Witten diagrams.  We show by explicit computation that the tree-level four-point and one-loop two-point functions agree with those computed using the standard Schwarzian action coupled to boundary bilocal operators. It is illuminating to see how the Schwarzian action naturally materializes in our approach based on a purely scalar field action.    

In order to solidify the connection to the Schwarzian approach, we go further and show that the two formulations give the same results for all tree-level  AdS$_2$ boundary correlators. At loop level the story is less clear due to UV issues. Note that above we claimed agreement for the one-loop two-point function; more precisely, this holds in the simplest renormalization scheme in which one sets to zero scale-free divergent integrals.   

The late-time correlation functions on $\Ic^+$ that we obtain are thus closely related to thermal AdS$_2$ correlators, as expected on general grounds.  They have a nice physical interpretation in terms of corrections to leading-order black hole thermodynamics. For example, the one-loop correction to the energy flux can be accounted for by a shift of the black hole mass at fixed temperature, yielding the corrected mass–temperature relation 
\eq{z3aa}{ M=\frac{Q}{\sqrt{G}}+2 \pi^2 \sqrt{G} Q^3 T_H^2+\frac{3}{2} T_H+\ldots \quad \Rightarrow \quad   \ln Z=\ln Z_{\mathrm{cl}}+\frac{3}{2} \ln T_H~.}
This reproduces the one-loop Schwarzian result of \cite{Stanford:2017thb}, here obtained from a Lorentzian computation of scalar field correlators, as opposed to a Euclidean path integral. Other than computational complexity, there is nothing to stop us from computing correlation functions at earlier times as well, where the results will depend on details of the collapse process. 

The main result of this work is to show how the onset of large quantum-gravity effects can be derived within a relatively conventional field-theoretic framework that makes minimal assumptions and connects smoothly to Hawking’s original treatment.  Our results hold in the temperature window
\eq{z2aa}{ {G\over r_0^3} \ll T_H \ll  {1\over r_0}~.}
The lower bound comes from requiring $g\ll1 $, while the upper bound defines the near-extremal regime. 

The rest of this paper is organized as follows.  In section \rff{RNsol} we review the \RN solution, its near-extremal, near-horizon scaling limit, and its formation by collapse of a charged spherical shell.    The Hamiltonian formulation is worked out in section \rff{Ham}, with the output being the gravitationally dressed scalar field action in both the full geometry and in the near-horizon region.  In section \rff{AdS2corrs} we use this action to compute some boundary correlators in the near-horizon AdS$_2$ region at tree-level and one-loop. These agree with results in the Schwarzian formulation, and indeed in the course of our computation we ``rediscover" the Schwarzian action.    In section \rff{equiv}   we go further, and demonstrate the equivalence, at the level of all tree-level correlators, between our Hamiltonian approach and the Schwarzian plus dressed bilocal operator approach.    The main results, correlators of the Hawking radiation at future null infinity, are presented in section \rff{results}, where we also give physical interpretations of the gravitationally induced corrections.   We close with a discussion in section \rff{discuss}, where we compare to the Schwarzian approach, and comment on some possible extensions for future study. A number of Appendices contain technical details and additional applications of our approach.

\section{\RN solution}
\label{RNsol}

In this section we review the standard \RN metric, its near-extremal limit, and its formation by the collapse of an infalling spherical, charged,  null shell. 

\subsection{Metric}

The  $d=3+1$ \RN black hole with mass $M$ and charge $Q$ has the line element
\eq{b1}{  ds^2 = -f(r)dt^2 + {dr^2\over f(r)}+ r^2 d\Omega^2~,}
with
\eq{b2}{ f(r)&= 1-{2GM\over r } +{GQ^2\over r^2} = {(r-r_+)(r-r_-) \over r^2}
\cr  r_\pm &= GM \pm \sqrt{G^2M^2 -GQ^2}~. }
  The event horizon is located at $r=r_+$.   The Hawking temperature is
\eq{b3}{ T_H = {r_+ - r_- \over 4\pi r_+^2}~. }

\subsection{Near-extremal limit}
\label{near1}

The extremal limit is obtained as  $r_+ \rt  r_-$ or equivalently  $\sqrt{G}M \rt  |Q|$; we henceforth take $Q>0$. Of relevance here will be the near-extremal, near-horizon scaling limit. This is obtained by writing 
\eq{b4}{  r_\pm & = r_0 \pm \lambda \rho_h \,, }
and taking $\lambda \rt 0$ at fixed $r_0$ and $\rho_h$, though we will be interested in small but finite $\lambda$. To exhibit the near-horizon geometry we simultaneously change coordinates $(r,t) \rt (\rho,\ttil)$ as 
\eq{b5}{ r& = r_0 +\lambda \rho~,\cr 
t & = {r_0^2 \over \rho_h \lambda} \ttil \,.}
Taking $\lambda \rt 0$ at fixed values of the new coordinates gives 
\eq{b6}{ f =  {\rho^2 - \rho_h^2\over r_0^2}\lambda^2 +  \ldots  \,,}
and
\eq{b7}{ ds^2 = r_0^2 \left( -\ft d\ttil^2+ {d\rho^2 \over \rho_h^2\ft}  + d\Omega^2 \right) + \ldots \,,    }
with
\eq{b8}{ \ft = {\rho^2 -\rho_h^2\over \rho_h^2}~.}
Equation \rf{b7} is the familiar AdS$_2 \times S^2$ near-horizon geometry, where the $\ldots$ encode the deviations that (for finite $\lambda$) incorporate the flat asymptotics as $r\rt \infty$. 

It will also be convenient to trade the radial coordinate $\rho$ for the dimensionless coordinate $z$ via
\eq{b9}{ \rho = \rho_h \coth z\,, }
in terms of which 
\eq{b10}{ ds^2 = r_0^2\left(  {-d\ttil^2+ dz^2 \over \sinh^2 z} + d\Omega^2 \right) + \ldots ~.}
 In these coordinates, the event horizon is at $z=\infty$ while the AdS$_2$ boundary is approached when $z\rt 0$. We also note the following behavior of the Hawking temperature
\eq{b10a}{ T_H = {r_+ -r_- \over 4\pi r_+^2}  \approx { \rho_h \over 2\pi  r^2_0} \lambda\,.}
In the near-extremal limit, the mass behaves as 
 \eq{b10b}{  M = M_{\rm ext}    + 2 \pi^2 \sqrt{G} Q^3 T_H^2+O\left(T_H^3\right)~,\quad M_{\rm ext} = {Q\over \sqrt{G}}~.}

\subsection{Collapse geometry}

We now turn to  the collapse solution shown in
Figure \rff{penrose}.   The resulting Vaidya geometry is given by Minkowski space and  \RN   matched across an infalling null shell. 
The Minkowski region is given by 
\eq{b11}{ ds^2 & = -dT^2+ dr^2+ r^2 d\Omega^2 \cr
&  =-dUdV + r^2(U,V)d\Omega^2~.}
with 
\eq{b12}{  U = T-r~,\quad V=T+r\,,}
where the shell trajectory is given by $V=V_s$.    Outside the shell we introduce the light cone coordinates
\eq{b13}{ u=t-r_*~,\quad v=t+r_* \,,}
with $r_*$ given by solving   
\eq{b14}{  {dr_* \over dr } = {1\over f(r)} \,.}
The explicit solution is given by 
\eq{b15}{ r_* =   r+ {r_+^2 \over r_+-r_-} \ln(r-r_+) -  {r_-^2 \over r_+-r_-} \ln(r-r_-)~. }
The  \RN line element is then 
\eq{b16}{ ds^2 = f\big(r(r_*)\big)\left(-dt^2 + dr_*^2\right) + r^2(r_*)d\Omega^2~.}
The shell trajectory in the light cone coordinates is $v=v_s$.   

We now match the coordinates across the shell.  We are free to choose $V=v$ so that $V_s=v_s$. The relation between $U$ and $u$ is fixed by demanding $r(U)= r(u)$ along the shell trajectory, which amounts to imposing
\eq{b17}{  {v_s-U\over 2} = r\Big( r_*= {v_s-u\over 2}\Big)~.}
This gives
\eq{b18}{ u & = U-{2r_+^2 \over r_+-r_-} \ln\left({v_s-U\over 2}-r_+\right) +  {2r_-^2 \over r_+-r_-} \ln\left({v_s-U\over 2}-r_-\right)~. }
The event horizon at $u=\infty$ is mapped to the finite value
\eq{b19}{ U=U_h\equiv  v_s -2r_+~.}
The region $U> U_h$ is inside the horizon. 
Although it is not possible  to invert \rf{b18} analytically to find $U(u)$, the late time ($u\rt \infty)$ behavior can be obtained as
\eq{b20}{ U = U_h - C e^{-2\pi T_H u} + \ldots~,\quad C =2 (r_+-r_-)^{r_-^2/r_+^2} e^{2\pi T_HU_h}~.}

\section{Hamiltonian formulation of Einstein-Maxwell-scalar theory}
\label{Ham}

In this section we use the Hamiltonian formulation to work out a gauge-fixed action for the s-wave sector of a scalar field coupled to Einstein-Maxwell theory.   The result is an action for the scalar field, local in time but non-local in space, that captures the gravitational self-interactions of the scalar field.  Since we do not use the equations of motion to obtain this action, it is suitable for use in a path integral. 

\subsection{Action in Hamiltonian form}

We start from the action for $d=3+1$ Einstein-Maxwell theory coupled to a real scalar field, 
\eq{a1}{ S = - {1\over 4\pi}\int\! d^4x \sqrt{-g} \left[  {1\over 4G} {\cal R} +{1\over 4} F_{\mu\nu}^2  +{1\over 2} (\nabla \phi)^2 +V(\phi) \right]~.}
Boundary terms will be fixed later once we convert to the Hamiltonian formulation. We reduce to the s-wave by writing 
\eq{a2}{ ds^2 & = -N^2(t,r) dt^2+ L^2(t,r) \big(dr + N^r(t,r) dt\big)^2 + R^2(t,r)d\Omega^2 \cr
  A& = A_t(t,r)dt +A_r(t,r)dr \cr
  \phi& = \phi(t,r)~.}
As written, the action \rf{a1} makes no reference to the collapsing shell that forms the black hole. The idea is that we expect all of the interesting interactions to occur in the region outside the shell (since inside the shell we simply have Minkowski space) and so it is implicit that the integration region is restricted to the exterior of the shell. Alternatively, were we to consider scalar field interactions in the eternal black hole, we would restrict the integration region to lie outside the horizon. Both cases will be considered below.  Note that we are not directly quantizing the shell itself; using standard WKB-type arguments, we expect that any independent fluctuations of the shell are suppressed by its mass.

To carry out the Legendre transformation to the Hamiltonian form we use the relation 
\eq{a3a}{ {1\over 4\pi} \int_{S^2}d^2\Omega\sqrt{-g}\Rc & =-\frac{2L(N^r)^2(R')^2}{N}
-\frac{4LN^rRR'(N^r)'}{N}
-\frac{4(N^r)^2RR'L'}{N}
+\frac{4LN^rR'\dot R}{N}\cr
&\qquad
+\frac{4LR\dot R\,(N^r)'}{N} 
+\frac{4N^rRR'\dot L}{N}
+\frac{4N^rR\dot R\,L'}{N}
-\frac{2L\dot R^{\,2}}{N}
-\frac{4R\dot R\dot L}{N}   \cr
& \qquad
+\frac{2N(R')^2}{L}
+\frac{4N'RR'}{L} +2NL+ ({\rm tot~deriv})~. }
Here we have denoted time derivatives with a dot and spatial derivatives with a prime. Some algebra gives
\eq{a3}{ S = \int\! dt dr \big[ \pi_\phi \dot{\phi} +\pi_R \dot{R} +\pi_L \dot{L} -N( {\cal H}^\phi_t +\Hc^G_t)-N^r ( {\cal H}^\phi_r +\Hc^G_r) \big]  -\int \! dt H_{\rm ADM}\,,    }
with 
\eq{a5}{ \Hc^\phi_t &= {1\over 2}\left( {\pi_\phi^2 \over LR^2}+{R^2 \over L} {\phi'}^2\right)+R^2L V(\phi)~,\quad \Hc^\phi_r = \pi_\phi \phi' \\
\Hc^G_t & = {GL\pi_L^2 \over 2R^2}-{G\pi_L \pi_R \over R} +{1\over G}\left[ \left({RR'\over L}\right)'-{{R'}^2 \over 2L}-{L\over 2} \right]  + { Q^2 L \over 2R^2}~,\quad \Hc^G_r =R'\pi_R -L\pi'_L~.\notag \cr }
In \rf{a3} we have added the ADM boundary term $ -\int \! dt H_{\rm ADM}$, to be fixed below. 
Also,  we have taken the shortcut of solving Gauss' law and inserting the solution back into the action. In particular, the Gauss law constraint reads $(\pi_{A_r})'=0$, which we solved  as $\pi_{A_r}=Q$. 

The Euler-Lagrange equations of \rf{a1} evaluated on the s-wave ansatz are equivalent to the Euler-Lagrange equations of \rf{a3} for the phase space variables $(\phi,\pi_\phi,L,\pi_L, R,\pi_R)$, along with the initial value constraints 
\eq{a5a}{  {\cal H}^\phi_t +\Hc^G_t =0~,\quad  {\cal H}^\phi_r +\Hc^G_r=0~.}
We can put the constraints into a more useful form by defining the quasilocal mass ${\cal M}$,
\eq{a6}{ \Mc =  {G\pi_L^2 \over 2R} +{R\over 2G} \left[ 1- \left({R'\over L}\right)^2  \right] +{Q^2 \over 2R}~,}
in terms of which the constraints take the form 
\eq{a7}{  \Mc' & = {R'\over L } \Hc^\phi_t   +{G\pi_L \over RL}  \Hc^\phi_r   \cr
  & R'\pi_R -L\pi'_L =- \pi_\phi \phi' ~.}
We consider solutions with $\lim_{r\rt \infty} N =1$ and $\lim_{r\rt \infty} N^r =0$, so that $t$ measures proper time in the asymptotically flat region. This gives the following expression for the ADM mass,
\eq{a8}{ H_{ADM} = \lim_{r\rt \infty} \Mc~,}
as can be derived, following Regge-Teitelboim \cite{Regge:1974zd},  by demanding a good variational principle; see \cite{Kraus:2025efu} for more details in the present context.  

To reduce the theory to the physical degrees of freedom, we need to impose two gauge conditions and then solve the two constraints.  To illustrate the procedure, we first show how to recover the \RN solution. To this end, we fix the gauge by setting 
\eq{a9}{ R=r~,\quad \pi_L=0 \,,}
and set the scalar field to zero:  $\phi = \pi_\phi=0$.  The second constraint equation in \rf{a7} sets $\pi_R=0$ while the first  is readily solved to give  
\eq{a10}{  L^2 = {1\over 1 - {2GM\over r} + {G Q^2\over r^2}}~,}
where we wrote $\Mc = M = $ constant. The equations of motion further yield $N^2 = {1\over L^2}$ and $N^r=0$, thereby yielding the standard form of the \RN solution in Schwarzschild coordinates. 

\subsection{Solving the constraints}

We now turn to solving the constraints in the presence of a general scalar field. We again choose the gauge \rf{a9}. The second equation in \rf{a7} gives $\pi_R =-\pi_\phi \phi'$ while the first equation reads (we are using the shorthand $V(r) = V\big(\phi(r,t)\big)$) 
\eq{a11}{ \left({r \over L^2}\right)'= 1-{GQ^2 \over r^2}- {2Gh(r)\over L^2}-2G r^2 V(r)~,}
with
\eq{a12}{ h(r) = {1\over 2} \left( {\pi_\phi^2 \over r^2} +r^2 {\phi'}^2\right)~.}
To integrate \rf{a11} we need to impose a boundary condition. In the case that a shell is present, located at $r=r_s$, since we are ignoring scalar field backreaction inside the shell, we should demand that $L$ take the same value right outside the shell as in the \RN solution.  Thus we impose  ${1\over L^2(r_s)}= f(r_s)$  where as  usual
\eq{a14a}{ f(r) = 1-{2GM\over r}+{GQ^2\over r^2}~.}
The corresponding solution is
\eq{a13}{ {1\over L^2(r)} & = f(r) -{2G\over r} \int_{r_s}^r \! dr'\Big( f(r') h(r')+{r'}^2 V(r')\Big)  e^{-\int_{r'}^r \! dr'' {2Gh(r'')\over r''}}~.  }
In the absence of the shell (i.e. for an eternal black hole) we replace the shell location by the horizon, so that the boundary condition is ${1\over L^2(r_+)}=0$.  Equation \rf{a13} then holds with the substitution $r_s \rt r_+$. 

The quasilocal mass is 
\eq{a14}{ \Mc(r) = {r\over 2G} \left(1-{1\over L^2(r)}\right) +{Q^2\over 2r}~, }
yielding the ADM Hamiltonian
\eq{a15}{ H_{\rm ADM} & = \Mc(\infty)=    M + \int_{r_s}^\infty \! dr'\Big( f(r') h(r')+{r'}^2 V(r')\Big)   e^{-\int_{r'}^\infty \! dr'' {2Gh(r'')\over r''}}~,  }
where we  normalize $t$ such that  $\lim_{r\rt \infty} N^2(r)=1$ so that $H_{\rm ADM}$ generates translations in $t$.  The reduced scalar field action (restricted to the region outside the shell) is then 
\eq{a16}{ S  = \int\! dt \int_{r_s}^\infty\! dr  \left( \pi_\phi \dot{\phi} -\Big(f(r)h(r) + r^2V(r)\Big)  e^{-\int_{r}^\infty \! dr' {2 Gh(r') \over r'} }   \right)\,. }
This reduced action for the scalar field is local in $t$ but nonlocal in $r$.  It captures all of the gravitational backreaction effects of the scalar field on the metric. We also emphasize that since we have only used the constraints and not the equations of motion, the resulting action is formally exact at the quantum mechanical level in the sense that it can be used to formulate a path integral.   We say ``formally" because we expect UV divergences and because the non-polynomial nature of the action requires special care. 

\subsection{Integrating out  the canonical momentum}

The Lagrangian form of the reduced scalar field action is obtained from \rf{a16} by integrating out the canonical momentum $\pi_\phi$. In general, this should be done at the level of the path integral; this will be discussed once we take the near-horizon limit, while here we consider the classical action at first order in the coupling simply for illustration.    

If we are interested in the on-shell action at first order in $G$, it suffices to compute $\pi_\phi$ to zeroth order and substitute back in. At this order, 
\eq{a17}{  \pi_\phi = {r^2  \over f(r)} \dot{\phi} + O(G) \,,}
so that 
\eq{a18}{ S  & =  \int\! dt \int_{r_s}^\infty\! dr  r^2 \left(  {1\over 2}  { \dot{\phi}^2\over f(r)} - {1\over 2} f(r) {\phi'}^2 -V(r) \right) \cr
& \quad +2G   \int\! dt \int_{r_s}^\infty\! dr \int_r^\infty\! dr' \Big( f(r)h(r) +r^2V(r)   \Big) {h(r') \over r'} + O(G^2) \,,}
where now 
\eq{a19}{ h(r) = {r^2 \over 2f(r) }  \left(   {\dot{\phi}^2\over f(r)}  + f(r)  {\phi'}^2 \right)\,. }
The first line of \rf{a18} is the usual free scalar action in the \RN metric $ds^2= -f(r)dt^2 + {dr^2\over f(r) }+ r^2 d\Omega^2$, and the second line gives the leading gravitational self-interaction correction.

 \subsection{Near-horizon limit}
 
 We now wish to apply the near-extremal, near-horizon limit to the action \rf{a16}. Along with the scaling in section \rff{near1}, we need to scale $(\phi,\pi_\phi)$.   Collecting all the expressions we have
 \eq{c1}{  r_\pm & = r_0 \pm \lambda \rho_h\,, \cr
 t & =  {r_0^2 \over \rho_h \lambda} \ttil~,\quad 
 r = r_0 + \lambda \rho_h \coth z\,, \cr
 \phi &= {1\over r_0}\tilde{\phi}~,\quad 
 \pi_\phi  ={r_0\over \rho_h \lambda }\sinh^2z ~\tilde{\pi}_\phi~. }
 For definiteness, we now take 
 \eq{c2}{  V(\phi) = {1\over 2}m^2 \phi^2\,,}
 and also define   
 \eq{c2a}{ \tilde{m} = r_0 m~.}
 Applying the scalings to \rf{a16} and taking $\lambda \rt 0$ gives  
\eq{c3}{ S = \int\! d\tilde{t} \int_0^{z_s} \! dz\Big[  \tilde{\pi}_\phi \dot{\phit} -{1\over 2}  \Big(  {\tilde{\pi}_\phi}^2  + {{\phit}'}{}^2    +   {\tilde{m}^2\over \sinh^2 z}  { \phit}^2   \Big)  e^{- {1\over 2} g \int_0^z \! dz' \sinh^2 z'  (  {\tilde{\pi}_\phi}^2   + {\tilde{\phi}}'{}^2   )  }  \Big] \,. }
Here we have defined the effective coupling 
\eq{yy15}{ g = {2G \over \rho_h r_0 \lambda} \approx {G \over \pi r_0^3 T_H } \,. }
where we recall the near-extremal expression $T_H = {r_+ -r_- \over 4\pi r_+^2}  \approx { \rho_h \over 2\pi  r^2_0} \lambda$. Note that with our definitions all quantities appearing in $S$ are dimensionless. The growth of the effective coupling $g$ as $T_H\rt 0$ encodes the appearance of strong coupling effects in the near-horizon region, invalidating the free field approximation as we approach extremality.
 
The action \rf{c3} may be used to compute correlation functions of the scalar field, including those at the boundary of the AdS$_2$ region. In section \rff{equiv}, we show that at tree level all such boundary correlators agree with those computed from the alternate JT gravity plus scalar bilocal operator description. At loop level the correlators are  UV sensitive, as we first show in section \rff{intout}. One distinction worth noting is that our action is fully gauge fixed, while in the JT plus bilocal description there is a residual SL(2,$R$)  gauge redundancy.  
  
\section{Perturbative scalar correlators in \texorpdfstring{AdS$_2$}{AdS2}}
\label{AdS2corrs}
 
 In this section, we use the action \rf{c3} to compute certain AdS$_2$  boundary correlation functions to lowest order in the effective coupling $g$. After expanding the action to the required order, we simply compute the corresponding Witten diagrams. In particular, to compare with Maldacena-Stanford-Yang (MSY) \cite{Maldacena:2016upp}, we compute the tree-level four-point function and one-loop two-point function in Euclidean signature.   The agreement with results in MSY \cite{Maldacena:2016upp} provides a useful check of our general methodology.  Furthermore, we will use these results later to obtain correlation functions at null infinity. 
 
 To reduce clutter, in this section we drop the tildes and write the action as 
\eq{d1}{ S = \int\! dt \int_0^\infty \! dz\Big[  {\pi}_\phi \dot{\phi} -{1\over 2}  \Big(  \pi_\phi^2  + {{\phi'}}^2    +   {{m}^2\over \sinh^2 z}  { \phi}^2   \Big)  e^{- {1\over 2} g \int_0^z \! dz' \sinh^2 z'  (  \pi_\phi^2   + {{\phi}'}^2)  }  \Big]   }
where we set $z_s=\infty$ corresponding to an eternal black hole. 
We further simplify by restricting to the massless case,
\eq{d1a}{ m=0~.}

\subsection{Integrating out \texorpdfstring{$\pi_\phi$}{pi_phi}}
\label{intout}

Using the flat measure associated with the symplectic form $\int\! dz \delta \pi_\phi \wedge \delta \phi$ the (formal) path integral is 
\eq{d2}{   Z = \int  [D\pi_\phi D \phi] e^{iS[\phi,\pi_\phi]}~.}
We aim to integrate out $\pi_\phi$ to the order in $g$ needed to compute the correlation functions mentioned above; in particular, we need the resulting Lagrangian action to first order in $g$.  
To this end, the action expanded to order $g$ is 
\eq{d3}{ S[\phi,\pi_\phi] & = \int\! dt \int_0^\infty \! dz \Big[  {\pi}_\phi \dot{\phi} -{1\over 2}  \Big(  \pi_\phi^2  + {{\phi'}}^2    \Big)   \Big]   \cr
& \quad   +{g\over 4} \int\! dt \int_0^\infty\! dz \int_0^z \! dz' \Big(  \pi^2_\phi(z)  + {\phi'}{}^2 (z)    \Big) \sinh^2 z'  \Big(  \pi^2_\phi(z')  + \phi'{}^2  (z')    \Big) \,.   }
Our task is therefore to compute 
\eq{d4}{ e^{iS[\phi]} &=  \int\! [D\pi_\phi] e^{i  \int\! dt \int_0^\infty \! dz \left[  {\pi}_\phi \dot{\phi} -{1\over 2}  \left(  \pi_\phi^2  + {{\phi'}}^2    \right)   \right]  }\cr
& \quad\quad\!\! \times   \left[  1 +{ig\over 4} \int\! dt \int_0^\infty\! dz \int_0^z \! dz' \Big(  \pi^2_\phi(z)  + {\phi'}{}^2 (z)    \Big) \sinh^2 z'  \Big(  \pi^2_\phi(z')  + \phi'{}^2  (z')    \Big)   \right]\,. }
Writing 
\eq{d5}{ \pi_\phi = \dot{\phi}+  \hat{\pi}_\phi \,,}
we have
\eq{d6}{  e^{iS[\phi]} &=  e^{iS_0[\phi]} \int\! [D\hat{\pi}_\phi] e^{-{i\over 2} \int\! dt \int_0^\infty\! dz \hat{\pi}^2_\phi} \left[  1+ ig \int\! dt \int_0^\infty\! dz  \int_0^z\! dz' \sinh^2 z'  T_{tt}(z) T_{tt}(z')      \right.   \cr
&  +  ig  \int\! dt \int_0^\infty\! dz  \int_0^z\! dz' \sinh^2 z'   \left(  \Big[  {1\over 2} T_{tt}(z) \pih_\phi^2(z')+{1\over 2} T_{tt}(z') \pih_\phi^2(z)+ \phid(z)\phid(z') \pih_\phi(z)\pih_\phi(z')   \Big] \right.  \cr
&   \quad\quad\quad\quad \quad\quad \quad  \quad \quad \quad \quad \quad\quad \quad \quad    + {1\over 4}   \pih_\phi^2(z)\pih_\phi^2(z')    \Big)   \Big]  \,, }
with 
\eq{d7}{ S_0[\phi] = {1\over 2}\int\! dt \int_0^\infty\! dz  (\dot{\phi}^2 - {\phi'}^2)~,}
and 
\eq{d8}{ T_{tt} = {1\over 2} (\dot{\phi}^2 + {\phi'}^2)~.}
We dropped terms with an odd power of $\pih_\phi$ since they integrate to zero by symmetry.   We can ignore the ${1\over 4}   \pih_\phi^2(z)\pih_\phi^2(z') $  term since it just contributes a $\phi$ independent normalization factor that can be absorbed into the measure.  
The terms in the second line of \rf{d6} yield UV divergent contributions that induce operators quadratic in $\phi$ at order $g$.   We refer to these collectively as ${\cal O}_2$.    Such operators do not contribute to the four-point function at order $g$ but can in principle contribute to the two-point function at order $g$.    The contractions of the two $\pi_\phi$ fields occur at equal time, and hence yield $\delta(0)$ factors. In momentum space this is a scale-free integral, set to zero in dimensional regularization.   Furthermore, we will see below that we get agreement with MSY \cite{Maldacena:2016upp} by simply dropping such contributions.   As a provisional rule, we therefore set ${\cal O}_2=0$. 

To summarize, at order $g$ we can use the effective action for $\phi$ 
\eq{d9}{ S[\phi] & =  {1\over 2}\int\! dt \int_0^\infty\! dz  (\dot{\phi}^2 - {\phi'}^2) + g\int\! dt \int_0^\infty\! dz  \int_0^z\! dz' \sinh^2 z'  T_{tt}(z) T_{tt}(z')~.}
Rotating to imaginary time, $t=i\tau$, we have the Euclidean action 
 \eq{d10}{ I[\phi] & =    {1\over 2}\int_0^{2\pi} \! d\tau \int_0^\infty\! dz  (\dot{\phi}^2 + {\phi'}^2) - g\int_0^{2\pi} \! d\tau  \int_0^\infty\! dz  \int_0^z\! dz' \sinh^2 z'  T_{\tau\tau}(z) T_{\tau\tau}(z') \,, }
 where now
 \eq{d11}{ T_{\tau\tau} = {1\over 2} (\dot{\phi}^2 - {\phi'}^2)~.}
As indicated, we have taken $\tau$ to be $2\pi$ periodic, consistent with the periodicity associated with the Hawking temperature in the rescaled coordinates. 

To compute boundary correlators, we  will need the massless scalar bulk-boundary propagator,
\eq{d12}{ K\left(z, \tau ; \tau_i\right)& =\frac{1}{2\pi} \frac{\sinh z}{\cosh z -\cos \left(\tau-\tau_i\right)}\cr
& = {1\over 2\pi} \sum_{n=-\infty}^\infty e^{-|n|z -in(\tau-\tau_i) }\,,} 
which obeys the free wave equation with boundary condition $\lim_{z\rt 0} K\left(z, \tau ; \tau_i\right) = \delta(\tau-\tau_i)$, as well as the bulk-bulk propagator
\eq{d13}{  G\left(\tau, z ; \tau^{\prime}, z^{\prime}\right)&=\frac{1}{4 \pi} \log \left[\frac{\cosh \left(z+z^{\prime}\right)-\cos ( \tau-\tau')}{\cosh \left(z-z^{\prime}\right)-\cos ( \tau-\tau')}\right]  \cr
& =  \frac{z_{<}}{2 \pi}+\frac{1}{\pi} \sum_{n=1}^{\infty} \frac{1}{n} \sinh \left(n z_{<}\right) e^{-n z_>} \cos \left(n\left(\tau-\tau^{\prime}\right)\right)\,, } 
where $z_< = {\rm min}(z,z')$. The bulk-bulk propagator obeys
\eq{d14}{  \nabla_x^2 G = -\delta^{(2)}(x-x') \,,}
where $\int\! d^2x \sqrt{g} \delta^{(2)}(x-x')  =1$.   

We also record the lowest order boundary two-point function\footnote{The usual normalization factor $2\Delta-d$ equals $1$ for the present case of $\Delta=d=1$.}
\eq{d14a}{ G_2^0(\tau_1,\tau_2) & =   \lim_{z\rt 0} z^{-1} K(z,\tau_1,\tau_2) = {1\over 4\pi} {1\over \sin^2 {\tau_{12}\over 2}}~.}

 \subsection{Four-point function}
 
Following MSY \cite{Maldacena:2016upp}, we can simplify a bit by replacing the $\phi$ field by a pair of massless fields $V$ and $W$ and compute the mixed four-point function $\langle VVWW\rangle$.  This way we get only a single diagram rather than three for $\langle \phi \phi \phi \phi\rangle$.  The relevant terms in the corresponding action are  
\eq{d15}{ 
 I &=  \int\! d\tau dz   {1\over 2}\Big( \dot{V}^2 + V'{}^2 +\dot{W}^2+W'{}^2 \Big)  \cr
 &  -  g \int\! d\tau \int_0^{\infty} dz  \int_0^z \! dz'  \sinh^2 z'  \Big(   T^V_{\tau\tau}(\tau,z)  T^W_{\tau\tau}(\tau,z') + T^W_{\tau\tau}(\tau,z)  T^V_{\tau\tau}(\tau,z')   \Big)~.     }
Proceeding as usual for a Witten diagram with a quartic interaction, we contract the stress tensors with the corresponding bulk-boundary propagators, yielding the basic object
\eq{d16}{   T_{\tau\tau}(\tau,z;\tau_i,\tau_j) & =  \p_\tau K(z,\tau;\tau_i)  \p_\tau K(z,\tau;\tau_j)  -  \p_z K(z,\tau;\tau_i)  \p_z K(z,\tau;\tau_j)   \cr
& =   -{1\over (2\pi)^2} \sum_{n_1,n_2} ( |n_1 n_2|+ n_1 n_2) e^{ -(|n_1|+|n_2|)z -i(n_1+n_2)\tau}  e^{in_1\tau_i + in_2 \tau_j} \,,         }
where the factor of ${1\over 2}$ in $T_{\tau\tau}$ was cancelled by the two choices for the contractions.  We consider 
\eq{d17}{ G_4(\tau_i) = \langle V(\tau_1) V(\tau_2) W(\tau_3)W(\tau_4) \rangle~. }
Using \rf{d16} this is 
\eq{d18}{ G_4(\tau_i) & = g \int\! d\tau \int_0^\infty\! dz \int_0^z \! dz' \sinh^2 z' \Big( T_{\tau\tau}(\tau,z;\tau_1,\tau_2) T_{\tau\tau}(\tau,z';\tau_3,\tau_4)\cr& \quad \quad \quad\quad \quad \quad \quad \quad  \quad \quad \quad \quad \quad \quad + T_{\tau\tau}(\tau,z;\tau_3,\tau_4) T_{\tau\tau}(\tau,z';\tau_1,\tau_2)   \Big)~.    }
The integrals are elementary, and we find
\eq{d19}{ G_4(\tau_i)& = {g\over (2\pi)^3} \sum_{n_1\ldots n_4} \Big( { |n_1 n_2|+n_1n_2 \over |n_1|+|n_2|  }  \Big) {2\over   (|n_1|+|n_2|+|n_3|+|n_4|)^2 -4}  \Big({ |n_3 n_4|+n_3n_4 \over |n_3|+|n_4|  }   \Big) \cr
& \quad \times e^{i n_1 \tau_1+i n_2 \tau_2+i n_3 \tau_3+i n_4 \tau_4}  \delta_{n_1+n_2+n_3+n_4,0}~. }
The sums receive contributions only when $n_1 n_2>0$ and $n_3 n_4>0$.  We define 
\eq{d20}{ n_1+n_2=k, \quad n_3+n_4=-k \,. }
and because $(n_1,n_2)$ have the same sign, as do $(n_3,n_4)$, 
\eq{d21}{  |n_1| + |n_2| = | k|~,\quad  |n_3| + |n_4| = | k| \,.}
We therefore have 
\eq{d22}{ G_4(\tau_i)& =  {2g\over (2\pi)^3}  \sum_{|k|>1} \frac{J_k^V\left(\tau_1, \tau_2\right) J_{-k}^W\left(\tau_3, \tau_4\right)}{k^2\left(k^2-1\right)}\,, }
with
\eq{d23}{ 
J_k^V\left(\tau_1, \tau_2\right) & = \sum_{n_1, n_2}^{\prime} n_1 n_2 e^{i n_1 \tau_1+i n_2 \tau_2} \delta_{n_1+n_2, k}\,, \cr
J_{-k}^W \left(\tau_3, \tau_4\right) & =\sum_{n_3, n_4}^{\prime} n_3 n_4 e^{i n_3 \tau_3+i n_4 \tau_4} \delta_{n_3+n_4,-k} \,,
}
where $\sum' $ means to sum over same-sign modes only.  

The form of \rf{d22} suggests an interpretation in terms of the exchange of a field with propagator  ${1\over k^2(k^2-1)}$.  This is precisely the Schwarzian propagator, and the absence of the $k=0,\pm 1$ modes, usually ascribed to SL(2,$R$) gauge symmetry, here follows automatically.  We will therefore integrate in the Schwarzian field in order to facilitate comparison to MSY \cite{Maldacena:2016upp}.  

We define the free Schwarzian action
\eq{d24}{ I_{\rm Sch} =  {2\pi \over g} \sum_{|k|>1} k^2(k^2-1) \eps_k\eps_{-k}~,}
and the Schwarzian-current coupling
\eq{d25}{ C(\tau_i,\tau_j) & = -{i\over \pi}  \sum_{|k|>1} {\rm sgn}(k) J_k(\tau_i,\tau_j) \eps_{-k}  }
with $J_k$ given by \rf{d23}.  By construction,  $G_4$ in \rf{d22}  is reproduced by 
\eq{d26}{  G_4(\tau_i) & =  \int [D\eps]  e^{-I_{\rm Sch} } C^V(\tau_1,\tau_2) C^W(\tau_3,\tau_4) }
where for convenience we have normalized the measure such that 
\eq{d26a}{\int [D\eps]  e^{-I_{\rm Sch} } =1~. }
On the other hand,  the MSY \cite{Maldacena:2016upp} four-point function takes the form 
\eq{d27}{ G_4^{\rm MSY}(\tau_i) = {1\over\pi^2}  \int [D\eps]  e^{-I_{\rm Sch} } {B^V(\tau_1,\tau_2) B^W(\tau_3,\tau_4) \over  \big( 2\sin {\tau_{12} \over 2} \big)^2 \big( 2\sin {\tau_{34} \over 2} \big)^2}   }
where $\tau_{ij}= \tau_i-\tau_j$, and
\eq{d28}{ B(\tau_i,\tau_j) =  \eps'(\tau_i) +\eps'(\tau_j)  -\big(\eps(\tau_i)-\eps(\tau_j)\big) \cot {\tau_{ij} \over 2}~. }
  $B(\tau_i,\tau_j)$ is the expansion to first order in $\eps$ of the general dressed bilocal operator with $\Delta=1$, as reviewed in Appendix \rff{corr}.      The equivalence of the two versions of the  four-point function then follows from the identity
  \eq{d29}{    C(\tau_i,\tau_j) = {1\over \pi} {B(\tau_i,\tau_j) \over   \big( 2\sin {\tau_{ij} \over 2} \big)^2}\,, }
 as can be shown by a Fourier expansion of the right-hand side; see Appendix \rff{corr}.  
 
 The functional form of the $\tau$-space four-point function depends on the ordering of the $\tau_i$.   In principle one can compute the sum \rf{d22}, but it is simpler to pass to the Schwarzian plus bilocal description in position space and then Wick contract.   Taking $2\pi > \tau_1 > \tau_2 > \tau_3 > \tau_4 > 0$  the result is 
 %
 \eq{d30}{ G_4(\tau_i)  = {g\over 4\pi^3}{  \left( -2+ \tau_{12} \cot {\tau_{12} \over 2}  \right) \left( -2+ \tau_{34} \cot {\tau_{34} \over 2}  \right) \over  \big( 2 \sin {\tau_{12}\over 2}\big)^2  \big( 2 \sin {\tau_{34}\over 2}\big)^2 }~.}
 %
  
\subsection{Two-point function at one-loop}

We now return to the case of a single scalar field $\phi$ with the quartic interaction 
\eq{d31}{  I_4 =  -  g \int\! d\tau \int_0^{\infty} dz    \int_0^z \! dz'  \sinh^2 z'   T_{\tau\tau}(\tau,z) T_{\tau\tau}(\tau,z')     }
  with $T_{\tau\tau} =  {1\over 2} (\phid{}^2 - {\phi'}{}^2)$.   
  
There are two distinct diagrams, as shown in Figure \rff{2point}, one of which involves a contraction of fields in a single stress tensor. 
\begin{figure}[H]
\centering

\scalebox{0.7}{%
\begin{tikzpicture}[
    line cap=round,
    line join=round,
    x=1cm,
    y=1cm
]

\begin{scope}[shift={(0,0)}]
    \draw[thick] (0,0) circle[radius=1.65];

    \coordinate (Va) at (-0.08,0.15);
    \draw[thick] (Va) -- (75:1.65);
    \draw[thick] (Va) -- (22:1.65);

    \coordinate (Pa) at (0.13,0.73);
    \coordinate (Qa) at (0.79,0.40);

    \draw[
        thick,
        decorate,
        decoration={
            snake,
            amplitude=1.0mm,
            segment length=3.0mm
        }
    ] (Pa) -- (Qa);

    \fill (Pa) circle[radius=1.8pt];
    \fill (Qa) circle[radius=1.8pt];

    \node[font=\large] at (0,-2.08) {(a)};
\end{scope}

\begin{scope}[shift={(4.4,0)}]
    \draw[thick] (0,0) circle[radius=1.65];

    \coordinate (Vb) at (-0.05,0.12);
    \draw[thick] (Vb) -- (68:1.65);
    \draw[thick] (Vb) -- (25:1.65);

    \fill (Vb) circle[radius=1.8pt];

    \draw[thick] (-0.83,-0.43) circle[radius=0.42];

    \coordinate (Pb) at (-0.53,-0.13);
    \fill (Pb) circle[radius=1.8pt];

    \draw[
        thick,
        decorate,
        decoration={
            snake,
            amplitude=1.0mm,
            segment length=3.0mm
        }
    ] (Pb) -- (Vb);

    \node[font=\large] at (0,-2.08) {(b)};
\end{scope}

\end{tikzpicture}%
}

\caption{Diagrams for the one-loop two-point function, with quartic vertices pictured in terms of graviton exchange.  Diagram (b) involves a self-contraction of fields within a single stress tensor, which will turn out to vanish.  }
\label{2point}
\end{figure}
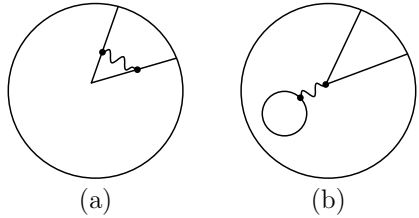

  
 The contractions for diagram (a)   amount to the replacement
 \eq{d32}{  T_{\tau\tau}(\tau,z) T_{\tau\tau}(\tau,z')   &~\rt ~   \p_\tau K_1\p_{\tau'} K_2 \p_\tau \p_{\tau'} G  \cr
&~  +\p_z K_1\p_{z'} K_2 \p_z \p_{z'} G   \cr
& ~ - \p_\tau K_1\p_{z'} K_2 \p_\tau \p_{z'} G   \cr 
& ~ -\p_z K_1\p_{\tau'} K_2 \p_z \p_{\tau'} G \cr
 &~  + \tau_1 \leftrightarrow \tau_2~. }
  Substituting in the Fourier expansions of the propagators and doing the integrals, which are elementary, gives
\eq{d33}{ G_2^{(a)} & =   -{g\over 4\pi^2}  \sum_{n=1}^\infty \sum_{n_1=1}^\infty { n n_1^2 \over \big( (n+n_1)^2-1\big) (n+n_1)^2}  \Big[ e^{i \tau_{12} n_1} + e^{-i \tau_{12} n_1}   \Big]~.  }

Next we consider diagram (b), which involves a self-contraction of the stress tensor.  The self-contraction of the stress tensor is fixed by the Weyl anomaly as 
\eq{d34}{  \langle T_{\mu\nu} \rangle& = \alpha g_{\mu\nu}  }
for some constant $\alpha$ whose value will not be needed.  So 
\eq{d35}{ \langle T_{\tau\tau}(\tau,z) \rangle = {\alpha \over \sinh^2 z}~,}
which is in particular independent of $\tau$.
 For the stress tensor that contracts with the bulk-boundary propagators, we have 
 \eq{d36}{  T_{\tau\tau}(\tau,z) & ~\rt ~  \p_\tau K(\tau,z;\tau_1) \p_\tau K(\tau,z;\tau_2) -  \p_z K(\tau,z;\tau_1) \p_z K(\tau,z;\tau_2) \cr
 & = - {1\over 2\pi^2} \sum_{n_1,n_2} \big( n_1 n_2 +|n_1n_2| \big) e^{-(|n_1|+|n_2|)z-i(n_1+n_2)\tau +in_1 \tau_1 +in_2\tau_2}~. }
Integration over $\tau$ forces $n_1=-n_2$ so that $n_1n_2 + |n_1 n_2|=0$. Therefore $G_2^{(b)}=0$, and the full two-point function is given by  \rf{d33}.   

On the other hand, the MSY \cite{Maldacena:2016upp} two-point function is given by 
\eq{d37}{ G_2^{\rm MSY} = \int\! [D\eps] e^{-I_{\rm Sch}} \left. {1\over \pi} \frac{T^{\prime}\left(\tau_1\right) T^{\prime}\left(\tau_2\right)}{\left[2 \sin \frac{T\left(\tau_1\right)-T\left(\tau_2\right)}{2}\right]^2} \right|_{\eps^2}\,,\quad \text{with}\quad T(\tau) = \tau+ \eps(\tau) \,, }
where, as indicated, we should expand the integrand to second order in $\eps$  and contract using the free Schwarzian action \rf{d24}.      Upon Fourier expanding this agrees precisely with \rf{d33}; see Appendix \rff{corr} for details.    The result in position space is 
\eq{d38}{ G^{1-{\rm loop}}_2 & =  {g\over 4\pi^2} {1\over \big(2 \sin {\tau\over 2} \big)^2} \Bigg[ \frac{\left(\tau^2-2 \pi \tau+2-2 \cos \tau+2(\pi-\tau) \sin \tau\right)}{4 \sin ^2 \frac{\tau}{2}}\cr
& \quad\quad\quad\quad\quad\quad\quad\quad +\frac{1}{2}\left(-2+\frac{\tau}{\tan \frac{\tau}{2}}\right)\left(-2+\frac{(\tau-2 \pi)}{\tan \frac{\tau}{2}}\right) \Bigg] }
with $\tau = \tau_1-\tau_2$ and we assumed $2\pi > \tau >0$. 

We now make some comments on the agreement.   First, we found perfect agreement without needing any contribution from the operator ${\cal O}_2$; as discussed above, ${\cal O}_2 =0$ in a renormalization scheme like dimensional regularization. Second, in the Hamiltonian computation we saw that only the diagram with no self-contractions contributed.    On the other hand, to obtain \rf{d38} from  \rf{d37}  one needs two distinct diagrams, one of which involves a contraction of two $\eps$ fields at the same location.  So the agreement between the two approaches is not diagram by diagram. 

\section{Classical equivalence of Hamiltonian and JT formulations}
\label{equiv}

The reduced Hamiltonian approach and the JT gravity plus dressed bilocal approach represent two different methods of fixing the gauge and handling the constraints.  As such, one expects agreement for physical quantities, at least at tree level.  It is, however, very instructive to demonstrate this explicitly; indeed, the equivalence is not completely obvious given the subtleties associated with boundary terms. This equivalence validates the Hamiltonian approach; however, we note that the detailed derivation that follows is not needed for the rest of this paper and so could be skipped by the reader willing to accept this claim.

We will take equivalence to mean equality of all tree-level scalar boundary correlators at noncoincident points.  As usual, the generating functional for these correlators is given by the on-shell action as a functional of the asymptotic boundary conditions, $S[\chi(t)]$.  For a massive scalar field associated with scaling dimension $\Delta$  (meaning $m^2 = \Delta(\Delta-1)$)  the asymptotic behavior is  
 \eq{e1}{   \phi(r,t) \approx r^{\Delta-1}\chi(t) + r^{-\Delta} \phi_\Delta(t) }
where we now write the AdS$_2$ metric in the form $ds^2 = - (r^2-1)dt^2 + {dr^2\over r^2-1}$, and we have set the AdS$_2$ radius to unity and taken the horizon to be at $r=1$.  In these coordinates the reduced scalar action is 
\eq{e2}{  S = \int\! d{t} \int_1^\infty d{r}  \Big[  \pi_\phi \dot{\phi} - \Big( {f}({r}) {h}({r}) +{V}({r})\Big) e^{-g \int_{{r}}^\infty d{r}' {h}({r}')} \Big]\,,  }
with
\eq{e3}{ {h} = {1\over 2}( \pi_\phi^2 + \phi'{}^2)~,\quad {f}  = {r}^2-1~,\quad V = {1\over 2} \Delta(\Delta-1) \phi^2~. }

To establish equivalence, we will show that the on-shell action obtained from \rf{e2} and that in the Schwarzian approach share the same differential expression ${\delta S[\chi] \over \delta \chi(\tau)}$.  We will also establish some conceptually illuminating results along the way.    

\subsection{Appearance of \texorpdfstring{AdS$_2$}{AdS2}}

The first step will be to rewrite the equations of motion of the action \rf{e2} as the wave equation for a massive scalar field on an AdS$_2$ metric whose coordinates themselves depend on the scalar field profile.   The Euler-Lagrange equations derived from \rf{e2} are 
\eq{e4}{  \dot{\phi} & =    F \pi_\phi~,\quad 
\dot{\pi}_\phi  =  (F\phi')' -EV_\phi \,,     }
where we have defined 
\eq{e5}{ E(r)&=  e^{-g \int_{{r}}^\infty d{r}' {h}({r}')}\,, \cr
F(r) &= f(r) E(r) -g \int_1^r \! dr' \Big( f(r') h(r') + V(r')\Big) E(r')\,, }
and $V_\phi = {dV\over d\phi}$. Here the function $F$ obeys the useful identity
\eq{e6}{ F' = E(f'-gV)~.}
Now define $L$ and $N$ as
\eq{e7}{ L & = \sqrt{E\over F}~,\quad N= FL=\sqrt{EF}~.}
This expression for $L$ is consistent with  \rf{a13} in the AdS$_2$ region. 
The Hamiltonian equations \rf{e4} become
\eq{e5a}{ \dot{\phi} = {N\over L} \pi_\phi~,\quad \dot{\pi}_\phi = \left({N\over L} \phi'\right)' -NL V_\phi~. }
These are the same equations as obtained for a scalar with potential $V$ in the metric 
\eq{e6a}{ ds^2 = - N^2 dt^2+L^2 dr^2~. }
It is worth noting that this result holds independent of any near-horizon approximation; i.e., it holds for the action \rf{e2} with $f(r)$ and $h(r)$ taking their full asymptotically flat forms. 

Specializing now to 
\rf{e3}, we will show that  the metric \rf{e6a} is AdS$_2$, namely that it obeys $\Rc=-2$. This condition is equivalent to 
\eq{e7a}{ -\frac{2}{N L}\left[\left(\frac{N^{\prime}}{L}\right)^{\prime}-\partial_t\left(\frac{\dot{L}}{N}\right)\right] =-2~. }
It remains to establish \rf{e7a}. 

Define 
\eq{e8}{  p = \pi_\phi \phi'~. }
Using the equations of motion, we deduce 
\eq{e9}{ \dot{h}& = Fp'+2 F'p -EV_\phi \pi_\phi    \cr
\dot{p} & = Fh'+2F'h-EV_\phi \phi'~.    }
Next define
\eq{e10}{ Y = {1\over L^2}~,}
which is seen to obey
\eq{e11}{  \dot{Y}' = -gV_\phi \dot{\phi} -g\dot{h}Y - gh\dot{Y}~.  }
Further define 
\eq{e12}{ X = \dot{Y} +g FpY \,,}
%
which, using identities derived above, obeys
\eq{e13}{ X' = -ghX~.}
Since $F(r=1)=0$, it follows that $Y(r=1)=0$ and therefore $X(r=1)=0$. Equation \rf{e13} then implies $X(r)=0$ for all $r$, so we have deduced 
\eq{e14}{ \dot{Y} = -g FpY \quad \Rightarrow \quad {\dot{L} \over N} = {1\over 2} gp~.  }

Next, using $f''=2$ we can show 
\eq{e15}{  \left({N'\over L}   \right)' =  {1\over 2} E \Big( 2+gYh'+2gh(f'-gV)-g V_\phi \phi' \Big)~.}
Using $E=NL$ and \rf{e9} we have
\eq{e16}{   \left({N'\over L}   \right)'   - NL = {1\over 2} g\dot{p}~.             }
From \rf{e14} this is 
\eq{e17}{   \left({N'\over L}   \right)'   - NL  = \p_t \left( {\dot{L}\over N}\right)\,, }
which is equivalent to \rf{e7a}, confirming that the metric is AdS$_2$. 

Let us summarize the content of these manipulations.  Given any, generally off-shell, trajectory through phase space corresponding to functions $\big(\phi(r,t),\pi_\phi(r,t)\big)$, we can define functions $(N,L)$ as in \rf{e7} and the corresponding metric $ds^2=-N^2 dt^2+ L^2 dr^2$.    Given these definitions, the Hamiltonian equations of motion for $(\phi,\pi_\phi)$ become the statement that the metric is AdS$_2$ and that $\phi$ obeys the wave equation for a free massive scalar on this AdS$_2$ metric.    Note that the AdS$_2$ nature of the metric only holds on-shell; off-shell, the metric defined here is fully determined by the scalar field variables but need not be constant curvature. 

\subsection{Appearance of the Schwarzian}

In the previous subsection, we established that on-shell we have a metric
\eq{e18}{  ds^2=-N(r,t)^2 dt^2+ L(r,t)^2 dr^2\,,}
which is AdS$_2$. The boundary is at some fixed large value of the radial coordinate, $r=r_b$. The metric functions have the following large $r$ asymptotics
\eq{e18a}{   N^2 & = r^2 +\alpha_N +\ldots \cr
{1\over L^2} & =r^2+ \alpha_L + \ldots \,.}
The fixed $r$ boundary has extrinsic curvature
\eq{e19}{ K = {N'\over NL} = 1- \left( \alpha_N -{1\over 2}\alpha_L\right){1\over r_b^2} +\ldots \,.}
Given that the metric is AdS$_2$, we can always define new coordinates $(Z,T)$ such that 
\eq{e20}{ ds^2 = {-dT^2 + dZ^2\over Z^2}~.   }
The fixed $r$ boundary now appears as a curve in the  $(Z,T)$ plane.  Defining a coordinate $u$\footnote{Not to be confused with the lightcone coordinate $u$ used elsewhere.} such that the induced line element on the boundary is $ds_{\rm bdy}^2=-r_b^2 du^2$ we write the boundary curve as $\big(Z(u),T(u)\big)$.   By equating the two forms of the induced metric on the boundary, we get for large $r_b$,
\eq{e21}{Z(u)  = {1\over r_b} T'(u)+ O(r_b^{-3})~.}
Starting from \rf{e20}, we want to carry out a change of coordinates to put the metric in the form \rf{e18} with the boundary at fixed $r$.  As a first guess, we can take the new coordinate to be $(r,u)$ corresponding to the coordinate transformation $Z = {1\over r} T'(u)$, $T=T(u)$.  This puts the boundary at fixed $r=r_b$  but gives an unwanted $dr du$ term in the metric. This cross term can be removed by setting 
\eq{e22}{  u = t+ {T''(t)\over 2T'(t) } {1\over r^2}~. }
Then we arrive at \rf{e18} with 
\eq{e23}{ N^2 &=  r^2  +{T'''(t)\over T'(t)}-2\left( {T''(t)\over T'(t)}\right)^2 + \ldots \cr
{1\over L^2}  & =  r^2 - \left( {T''(t)\over T'(t)}\right)^2  + \ldots  \,.}
The formula \rf{e19} then gives
\eq{e24}{ K= 1-{1\over r_b^2} {\rm Sch}\big(T(t),t\big)+\ldots   }
where 
\eq{e25}{ {\rm Sch}\big(T(t),t\big) =  {T'''(t)\over T'(t)}  -{3\over 2} \left( {T''(t)\over T'(t)}\right)^2~.}

\subsection{Relation between Hamiltonian and Schwarzian}

In the absence of boundary sources (i.e.,  for $\chi=0$), the Hamiltonian is given by 
\eq{e26}{ H_0 & = \int_1^\infty \! dr \Big( {f}({r}) {h}({r}) +{V}({r})\Big) e^{-g \int_{{r}}^\infty d{r}' {h}({r}')} \cr
& = {1\over g} \lim_{r\rt \infty} \left( f(r)-{1\over L^2(r)}\right)\cr
& =  -{1\over g} (1+\alpha_L)\cr
& =  {1\over g} \left( -1-2{\rm Sch}(T,t)+2(\alpha_N-\alpha_L)\right), }
where we used \rf{e19} and \rf{e24} to obtain the final line. In the above, we wrote $H_0$, because in the presence of a nonzero source $\chi$ the full Hamiltonian requires an additional boundary term to achieve a good variational principle, explicitly
\eq{e27}{  H= H_0 -(\Delta-1)\chi \phi_\Delta~.}
This is explained in MSY \cite{Maldacena:2016upp} and rederived in our language in Appendix \rff{variational}. There we also show that the full Hamiltonian has the on-shell variation
\eq{e28}{ \delta H = -(2\Delta-1)\phi_\Delta \delta \chi }
which identifies the boundary operator conjugate to $\chi$ as 
\eq{e29}{  \Oc_\Delta = (2\Delta-1)\phi_\Delta~.}

Next we compute $\alpha_N-\alpha_L$. On the one hand, this can be read off from 
\eq{e30}{ E(r) \approx 1 + {\alpha_N - \alpha_L \over 2r^2} +\ldots}
while on the other hand 
\eq{e31}{ \big( \ln E(r)\big)' = gh(r) \approx  -{g \Delta(\Delta-1)  \chi \phi_\Delta \over r^{3}}  }
where we used \rf{e1} and dropped $\chi^2$ terms since they do not contribute to correlators at separated points.  Putting these results together gives
 \eq{e32}{\alpha_N-\alpha_L =g \Delta(\Delta-1) \chi \phi_\Delta~. }
The full Hamiltonian is then found to be
\eq{e33}{ H = -{1\over g} -{2\over g} {\rm Sch}(T,t) +(\Delta-1) \chi \Oc_\Delta~.  }

It will be useful to compute the time derivative,
\eq{e34}{ {dH\over dt} = -{2\over g}{d\over dt}  {\rm Sch}(T,t) + {d\over dt} \left[ (\Delta-1) \chi \Oc_\Delta \right]~. }
On the other hand from  $\delta H = -\Oc_\Delta \delta \chi$  we have ${dH\over dt} = -  \Oc_\Delta \dot{\chi}$ and hence
\eq{e35}{ -\Oc_\Delta \dot{\chi} = -{2\over g}{d\over dt}  {\rm Sch}(T,t) + {d\over dt} \left[ (\Delta-1) \chi \Oc_\Delta \right]~. }
We also know
\eq{e36}{  \Oc_\Delta(t) = \frac{\delta S}{\delta \chi(t)}=2 D_{\Delta} T'(t)^{\Delta} \int \mathrm{d} t^{\prime} \frac{T'\left(t^{\prime}\right)^{\Delta} \chi\left(t^{\prime}\right)}{\left(T(t)-T\left(t^{\prime}\right)\right)^{2 \Delta}}~,\quad D_\Delta=\frac{\left(\Delta-\frac{1}{2}\right) \Gamma(\Delta)}{\sqrt{\pi} \Gamma\left(\Delta-\frac{1}{2}\right)}  }
as follows from solving the wave equation in the $(Z,T)$ coordinates and rewriting the result in terms of $t$.

\subsection{Computing the on-shell action, and agreement between the two formulations}

We now give an algorithm for computing the on-shell action as a functional of the boundary source $\chi(t)$. We start from the on-shell variation
\eq{e37}{  \delta S = \int\! dt \Oc_\Delta(t) \delta \chi(t)~.}
$\Oc_\Delta(t)$ is given in terms of $T(t)$ and $\chi(t)$ by \rf{e36}. An integro-differential equation for $T(t)$ is given by combining \rf{e35} and \rf{e36}, determining $T(t)$ in terms of $\chi(t)$, which in turn expresses $\Oc_\Delta(t)$ in terms of $\chi(t)$. Using this, \rf{e37} becomes a differential equation for $S$ as a functional of $\chi$.  

The same system of equations determines $S$ in the Schwarzian plus dressed bilocal operator description.  In this description, the action is 
\eq{e38}{ S = {2\over g} \int\! dt {\rm Sch}(T,t) + D_\Delta \int \mathrm{d} t_1 \mathrm{~d} t_2 \chi\left(t_1\right) \chi\left(t_2\right)\left[\frac{T'\left(t_1\right) T'\left(t_2\right)}{\left(T\left(t_1\right)-T\left(t_2\right)\right)^2}\right]^{\Delta}~. }
The operator expectation value is given by $ \Oc_\Delta = \frac{\delta S}{\delta \chi(t)}$ which yields the same expression as in \rf{e36}. Next we reproduce \rf{e35}.  Since this equation involves the time rate of change of energy, it can be obtained from Noether's theorem applied to the variation $\delta T(t) = \eta(t) T'(t)$.  Computing the corresponding variation of $S$ and setting it to zero (on-shell), we find precisely \rf{e35}.   Putting these statements together, we have reproduced the same system of equations as in the Hamiltonian formulation, establishing agreement between the respective on-shell actions.  This implies agreement of all tree-level boundary correlators.

\section{Correlations in Hawking radiation at null infinity}
\label{results}

We now turn to the main topic of interest: how to systematically compute correlations in the Hawking radiation produced by a black hole formed from collapse.   In particular, we are interested in the contributions that become large near extremality.    We first quickly review the free field theory case before turning to computing the corrections to the two-point and four-point functions at null infinity that exhibit the breakdown of the free field approximation near extremality.   We restrict attention to the case of a massless scalar field.  

\subsection{Hawking radiation in the free field limit}

The collapse geometry was reviewed in section \rff{RNsol}. We take the scalar field $\phi$ to be in the vacuum state in the far past, meaning on past null infinity ${\cal I}^-$. We denote this state by $|0\rangle$, and all expectation values will be computed in this state. An ingoing wave packet $f(v)$  localized inside the shell reflects off the origin $r=0$ to become an outgoing wave packet $f(U)$.   Positive frequency with respect to $v$ thus maps via the field equations to positive frequency with respect to $U$, which tells us that the outgoing part of the scalar field state right outside the shell is given by the $U$ vacuum.   

To describe the scalar field in the \RN portion of the geometry we  write 
\eq{f1}{ \phi(r,t) = {1\over r} \phit(r,t).}
Note that we previously used $\phit$ to denote the rescaled field in the near-horizon region, but from \rf{c1} we see that the  ${1\over r_0}$ rescaling used there agrees with \rf{f1} in the near-horizon region, so the definitions are compatible. In the \RN region the scalar field obeys the wave equation 
\eq{f2}{   \big(-\p_t^2 + \p_{r_*}^2 -V(r_*)\big) \phit=0 \,,}
where the potential is $ V(r_*) = {1\over r} {d^2 r \over dr_*^2}= { (r-r_+)(r-r_-)[r_+(r-r_-)+r_-(r-r_+)]\over r^6}$.   The potential vanishes as $r\rt r_+$ and as $r \rt  \infty$ and has a smooth bump in between.  The potential gives rise to nontrivial transmission and reflection coefficients of incident waves, corresponding to ``greybody factors."    We are interested in the near-extremal regime and  $\omega r_0 \ll 1$.   In this regime, an outgoing wave emanating from the horizon at $r_* =-\infty$ is reflected off the boundary of the near-horizon AdS$_2$ region with reflection coefficient ${\cal R}(\omega) = -1 +O(\omega r_0)$; see Appendix \rff{grey}.    This has important consequences for the quantum state in the near-horizon region.   In the low frequency regime, to good approximation we have a thermal Hartle-Hawking state in the near-horizon region, since the thermally populated outgoing modes reflect off the potential, yielding a thermal distribution of ingoing modes as well.  We will use this fact to relate near-horizon correlators in the collapse state to those in the thermal state for both ingoing and outgoing modes.  Outside the near-horizon region, the ingoing modes are in the vacuum state.  Finally, to propagate near-horizon correlators to future null infinity, we need to incorporate the small but nonzero transmission coefficient.  The precise form of the transmission coefficient is important for getting a numerically accurate result at future null infinity, but since it has nothing to do with the backreaction effects we are interested in, we will not dwell on this. In particular,  we proceed by first setting ${\cal T}(\omega)=1$ for the outgoing modes, restoring its actual value at the end of the computation.


With these comments in mind, we have the following free field correlators in the \RN part of the spacetime, outside the near-horizon region,
\eq{f3}{ \langle \p_U \phit(U_1,v_1) \p_U \phit(U_2,v_2)\rangle  & = -{1\over 4\pi} {1\over (U_1-U_2 -i\veps)^2}  \cr
 \langle \p_v \phit(U_1,v_1) \p_v \phit(U_2,v_2) \rangle  & =  -{1\over 4\pi} {1\over (v_1-v_2 -i\veps)^2}  \cr
  \langle \p_U \phit(U_1,v_1) \p_v \phit(U_2,v_2)\rangle  & =  0~.  }
The first line, which encodes the outgoing radiation, is of primary interest.  In the exterior region it is most convenient to work with the $u$ coordinate, in terms of which we have
\eq{f4}{  \langle \p_u \phit(u_1,v_1) \p_u \phit(u_2,v_2)\rangle= -{1\over 4\pi}{ \left( {\p u_1 \over \p U_1}  {\p u_2 \over \p U_2}\right)^{-1} \over (U_1-U_2 -i\veps)^2}~.  }
Recall that $u$ and $U$ are related as in \rf{b18}.  However, since \rf{b18} cannot be inverted analytically, we cannot obtain an explicit expression for \rf{f4} in terms of $u$. The exception is  at late times ($u \rt \infty$) where we can use \rf{b20} to obtain
\eq{f5}{ \langle \p_u \phit(u_1,v_1) \p_u \phit(u_2,v_2)\rangle &\approx -{\pi T_H^2\over 4\sinh^2 \big(\pi T_H(u_1-u_2-i\veps)\big)} ~, \quad  u\rt \infty~, }
which is a thermal correlator at temperature $T_H$, confirming the thermal nature of free field Hawking radiation at late times.  

We can also compute the stress tensor of the outgoing radiation.  Define
\eq{f6}{  \langle \Tt_{uu} \rangle = 2\pi \langle \p_u \phit \p_u \phit\rangle - {\rm counterterm} ~.}
$\tilde{T}_{\mu\nu}$ is the 2d stress tensor, while the 4d stress tensor contains a factor of ${1\over r^2}$ due to the rescaling \rf{f1}.  On $\Ic^+$, the counterterm is given by subtracting the stress tensor expectation value in the $u$ vacuum,
\eq{f7}{ \langle \Tt_{uu}(U_1) \rangle&  = -{1\over 2} \lim_{ U_2\rt U_1} \left[  { \left( {\p u_2 \over \p U_2}  {\p u_1 \over \p U_1}\right)^{-1} \over (U_2-U_1 -i\veps)^2}  - {1 \over (u_2-u_1-i\veps)^2} \right] \cr
& = {1\over 12} \left( {\p u_1 \over \p U_1}\right)^{-2} {\rm Sch} (u_1,U_1)  }
where the Schwarzian derivative is
\eq{f8}{  {\rm Sch} (u,U)   = {u'''(U)\over u'(U)}-{3\over 2} \left( {u''(U)\over u'(U)}\right)^2~. }
This is the standard result dictated by conformal invariance, and holds for any conformal field theory upon multiplying by the central charge $c$.   The explicit result in terms of  $U$ is
\eq{f9}{  \langle \Tt_{uu}(U) \rangle& = { 1 \over 6(2r_++U_h-U)^6}  \Big[  4r_+^2(r_+-r_-)^2 +12r_+(r_+-r_-)^2(U_h-U) \cr
& \quad\quad  +3(3r_++r_-)(r_+-r_-)(U_h-U)^2 +2(r_++r_-)(U_h-U)^3\Big] ~.}
The late time  ($u\rt \infty$ or $U\rt U_h$)  result is
\eq{f9a}{ \lim_{u\rt \infty}  \langle \Tt_{uu} \rangle =  {(r_+-r_-)^2 \over 96 r_+^4} =  {\pi^2 T_H^2 \over 6}~,}
which represents thermal radiation at temperature $T_H$.  The energy flux smoothly rises from zero at $u=-\infty$ to the late time thermal value \rf{f9a}. The physical energy flux measured at infinity falls off as $1/r^2$ due to the rescaling \rf{f1}.

For what follows, it will be useful to obtain \rf{f5} from a slightly different perspective.  We start with the near-horizon Euclidean two-point function \rf{d14a}, restoring the tildes on the coordinates,
\eq{f10}{ G_2^0(\tilde{\tau}_1,\tilde{\tau}_2) = {1\over 4\pi} {1\over \sin^2 {\tilde{\tau}_{12}\over 2}}~.}
This is the two-point function of the boundary operator $\Oc$, related to the bulk field as $\Oc =\lim_{\tilde{z}\rt 0}  \p_{\tilde{z}}\phit$, so \rf{f10} reads
\eq{f11}{ \lim_{z\rt 0}  \langle  \p_{z}\phit(z,\tilde{t}_1) \p_z\phit(z,\tilde{t}_2)\rangle_{\rm thermal} = - {1\over 4\pi} {1\over \sinh^2 {\tilde{t}_{12}-i\veps\over 2}}~.}
where we now rotated to Lorentzian signature and indicated that this is a thermal correlator.    As discussed above, to good accuracy we can take the state in the near-horizon region to be this thermal state.  Next, since $\tilde{u} = \tilde{t}+z$ and $\tilde{v}=\tilde{t}-z$ we have that $\p_{\tilde{u}}\phit = {1\over 2}(\p_z \phit +\p_{\tilde{t}} \phit)$.  The reflecting boundary condition at the near-horizon AdS$_2$ boundary (see Appendix \rff{grey}) implies $\p_{\tilde{t}} \phit =0$ at $z=0$, so we can use the relation $\p_{\tilde{u}}\phit= {1\over 2}\p_z \phit$.  Using this gives 
\eq{f12}{ \lim_{z\rt 0}  \langle \p_{\tilde{u}} \phit(\tilde{u}_1) \p_{\tilde{u}} \phit(\tilde{u}_2)\rangle = -{1\over 4}   {1\over 4\pi} {1\over \sinh^2 {\tilde{u}_{12}-i\varepsilon\over 2}}~.}
Finally, we convert to the original unscaled variables as 
\eq{f13}{ \tilde{u} = 2\pi T_H u }
to get 
\eq{f14}{ 
 \lim_{z\rt 0}  \langle \p_{u} \phit(u_1) \p_{u} \phit(u_2)\rangle = -   {\pi T_H^2 \over \big(2 \sinh \pi T_H (u_{12}-i\veps)\big)^2 }~.}
Since the correlator obeys the free field equation, it takes the form above for any $z$, and we thus get agreement with \rf{f5}. 

\subsection{Including interactions: general comments}

We now wish to include the effect of interactions.   We will treat the field inside the shell as free, and think of the initial state as being defined on the union of the shell trajectory with the $v>v_s$ portion of $\Ic^-$, as these together define a Cauchy surface. As discussed previously, the state on the shell is the $U$ vacuum, while on the portion of $\Ic^-$ we have the $v$ vacuum.  Our goal is to compute expectation values in this state of Heisenberg field operators inserted on $\Ic^+$.   This is a problem for in-in perturbation theory.

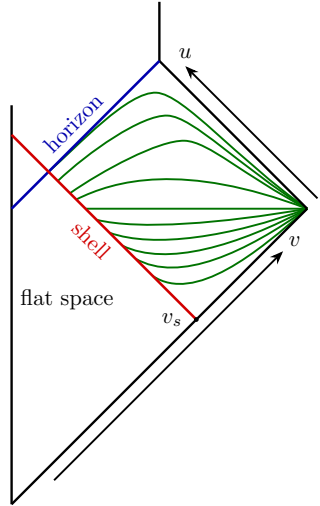
\begin{figure}[H]
\centering
\scalebox{0.85}{%
\begin{tikzpicture}[
    x=0.575cm,
    y=0.575cm,
    line cap=butt,
    line join=round,
    boundary/.style={black,line width=1.05pt},
    horizonline/.style={blue!68!black,line width=1.15pt},
    shellline/.style={red!82!black,line width=1.15pt},
    slice/.style={green!45!black,line width=0.85pt},
    axisarrow/.style={black,line width=0.85pt,-{Stealth[length=2.1mm,width=1.6mm]}},
    labelbox/.style={fill=white,fill opacity=0.92,text opacity=1,inner sep=1.3pt}
]

\useasboundingbox (-0.70,-0.45) rectangle (8.95,14.15);

\coordinate (Lbot) at (0,0);
\coordinate (Ltop) at (0,10.8);

\coordinate (Top) at (4,12);
\coordinate (TopUp) at (4,13.6);
\coordinate (R) at (8,8);

\coordinate (BlueStart) at (0,8);
\coordinate (ShellStart) at (0,10);
\coordinate (ShellEnd) at (5,5);

\draw[slice]
  (1.00,9.00)
  .. controls (1.82,9.72) and (3.15,11.18) .. (3.92,11.14)
  .. controls (4.66,11.10) and (6.10,9.34) .. (R);

\draw[slice]
  (1.25,8.75)
  .. controls (2.06,9.52) and (3.52,10.55) .. (4.22,10.52)
  .. controls (4.92,10.49) and (6.17,9.14) .. (R);

\draw[slice]
  (1.58,8.42)
  .. controls (2.44,9.02) and (3.98,9.86) .. (4.68,9.84)
  .. controls (5.34,9.82) and (6.27,8.90) .. (R);

\draw[slice]
  (1.75,8.25)
  .. controls (2.78,8.76) and (4.62,9.28) .. (R);

\draw[slice]
  (2.00,8.00) -- (R);

\draw[slice]
  (2.30,7.70)
  .. controls (3.45,7.42) and (5.70,7.60) .. (R);

\draw[slice]
  (2.65,7.35)
  .. controls (3.85,6.92) and (5.92,7.24) .. (R);

\draw[slice]
  (3.00,7.00)
  .. controls (4.20,6.42) and (6.12,7.00) .. (R);

\draw[slice]
  (3.35,6.65)
  .. controls (4.55,5.96) and (6.36,6.84) .. (R);

\draw[slice]
  (3.82,6.18)
  .. controls (4.96,5.42) and (6.66,6.82) .. (R);

\draw[boundary] (Lbot) -- (Ltop);
\draw[boundary] (Lbot) -- (R);
\draw[boundary] (Top) -- (R);
\draw[boundary] (Top) -- (TopUp);

\draw[horizonline] (BlueStart) -- (Top);
\draw[shellline] (ShellStart) -- (ShellEnd);

\path (BlueStart) -- (Top)
  node[
    pos=0.50,
    sloped,
    above=2.8pt,
    blue!68!black,
    font=\footnotesize,
    labelbox
  ] {horizon};

\path (ShellStart) -- (ShellEnd)
  node[
    pos=0.50,
    sloped,
    below=2.8pt,
    red!82!black,
    font=\footnotesize,
    labelbox
  ] {shell};

\fill[black] (ShellEnd) circle[radius=1.05pt];
\node[font=\footnotesize,left=2pt] at (ShellEnd) {$v_s$};

\draw[axisarrow] (8.28,8.28) -- (4.70,11.86);
\node[font=\footnotesize,above=2pt] at (4.70,11.70) {$u$};

\draw[axisarrow] (1.15,0.65) -- (7.35,6.85);

\node[
  font=\footnotesize,
  anchor=south west,
  xshift=0pt,
  yshift=-1pt
] at (7.25,6.8) {$v$};

\node[
    font=\footnotesize,
    fill=white,
    fill opacity=0.92,
    text opacity=1,
    inner sep=1.3pt
] at (1.50,5.55) {flat space};

\end{tikzpicture}%
}
\caption{The Hamiltonian is defined on curves of constant $t$, indicated by the green curves. We evolve the state over the region covered by these green curves.}
\label{fig:collapse-penrose}
\end{figure}
We employ our Hamiltonian formulation, with the Hamiltonian generating translations in Schwarzschild time outside the horizon.  Curves of constant time are indicated in green in figure \rff{fig:collapse-penrose}, and we need to evolve the Heisenberg fields over the region covered by the green curves. 

Working in the interaction picture, the Heisenberg fields obey (we suppress the spatial argument) 
\eq{f15}{ \phit(t) = U^\dagger(t,t_0)\phit_I(t) U(t,t_0) }
where 
\eq{f16}{  U(t,t_0) = T\left[ e^{-i\int_{t_0}^t \! dt' H'_I(t')}\right]~. }
Here, $t_0$ is an arbitrary reference time that we will take to coincide with the time at which the in state is defined. We have written the full  Hamiltonian as $H=H_0 + H'$, where $H_0$ is the free part.  $H'_I(t')$ denotes the interaction part with fields replaced by the interaction picture fields (which evolve under $H_0$).    

Consider a time-ordered four-point function
\eq{f17}{  G(t_i) & =  \langle \psi_{in}| \phit(t_1) \phit(t_2)\phit(t_3)\phit(t_4)   |\psi_{in}\rangle~,\quad  t_1>t_2>t_3>t_4>t_0~.}
Using \rf{f15}, this is rewritten as a correlator involving the free fields,
\eq{f18}{  G(t_i)  =  \langle \psi_{in}|U^\dagger(t_1,t_0)  \phit_I(t_1)U(t_1,t_2) \phit_I(t_2)U(t_2,t_3) \phit_I(t_3)U(t_3,t_4) \phit_I(t_4) U(t_4,t_0)  |\psi_{in}\rangle~.}
We then expand out the exponential $U$ operators to the desired order and Wick contract using the free two-point functions evaluated in $|\psi_{in}\rangle$.  Here we are using that $|\psi_{in}\rangle$ is a free field vacuum state, i.e., it is Gaussian.  The expression \rf{f18} can be drawn in terms of a Schwinger-Keldysh contour as shown in Figure \rff{SKa}, where the switchback segment corresponds to $U^\dagger(t_1,t_0) $.  
\newcommand{\skscale}{0.8}
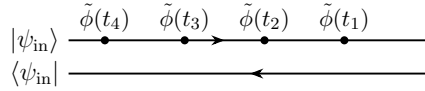
\begin{figure}[H]
\centering
\sbox0{%
\begin{tikzpicture}[
    line width=0.9pt,
    skdot/.style={circle,fill,inner sep=1.35pt},
    every label/.append style={font=\small, label distance=-2.5pt},
    skarrow/.style={
        postaction={decorate},
        decoration={markings, mark=at position #1 with {\arrow{Stealth[length=2.2mm]}}}},
    skarrow/.default=0.5,
]

\pgfmathsetmacro{\skpanelsep}{7.6}   
\pgfmathsetmacro{\sklen}{6.0}        
\pgfmathsetmacro{\skgap}{0.55}       
\pgfmathsetmacro{\skret}{2.2}        
\pgfmathsetmacro{\skvlen}{3.4}       

\pgfmathsetmacro{\skpA}{0.10}
\pgfmathsetmacro{\skpB}{0.32}
\pgfmathsetmacro{\skpC}{0.54}
\pgfmathsetmacro{\skpD}{0.76}

\begin{scope}[shift={(0,0)}]
  \draw[skarrow=0.43] (0,0) -- (\sklen,0);          
  \draw (\sklen,0) -- (\sklen,-\skgap);             
  \draw[skarrow=0.5] (\sklen,-\skgap) -- (0,-\skgap);

  \node[left] at (0,0)       {$\lvert\psi_{\rm in}\rangle$};
  \node[left] at (0,-\skgap) {$\langle\psi_{\rm in}\rvert$};

  \node[skdot,label={above:$\phit(t_4)$}] at ({\skpA*\sklen},0) {};
  \node[skdot,label={above:$\phit(t_3)$}] at ({\skpB*\sklen},0) {};
  \node[skdot,label={above:$\phit(t_2)$}] at ({\skpC*\sklen},0) {};
  \node[skdot,label={above:$\phit(t_1)$}] at ({\skpD*\sklen},0) {};
\end{scope}

\end{tikzpicture}%
}
\resizebox{\skscale\wd0}{!}{\usebox0}
\caption{Schwinger--Keldysh contours for the four-point function \rf{f17}.}
\label{SKa}
\end{figure}

In principle, \rf{f18} and its analog for other $n$-point functions may be used to compute correlation functions on $\Ic^+$, with the integration region being the region covered by the green curves in Figure \rff{fig:collapse-penrose}, covered twice due to the two segments of the Schwinger--Keldysh contour.   This is rather difficult to carry out.  However, as we have discussed, we are just interested in the enhanced near-horizon effects, and so we can restrict the integration to the near-horizon region depicted in Figure  \rff{fig:collapse-penrose-rcurve}.   Furthermore, since we assume free field behavior and incorporate the greybody factors at the end, we can first compute correlators on the outer boundary of the near-horizon region and then use the free field equation with vanishing potential to transport them directly to $\Ic^+$.

We have now reduced the problem to the much simpler one of computing a correlator at the boundary of an AdS$_2$ region cut off by the collapsing shell.   However, doing the integrals involving Lorentzian bulk-boundary propagators is still challenging, so we simplify further by focusing on late-time correlation functions.  At late times the outgoing part of the state appears thermal in free field theory, and due to the reflecting boundary condition at the boundary of the AdS$_2$ region this is also the case for ingoing modes in the near-horizon region. Thus we can proceed by computing a thermal AdS$_2$ correlator and then extract the relevant outgoing part using the same strategy as was used to obtain \rf{f14}.    Since we have already computed the needed thermal correlators, this is simple to execute.   The procedure is illustrated in Figure \rff{fig:SK}.
\begin{figure}[H]
\centering
\sbox0{%
\begin{tikzpicture}[
    line width=0.9pt,
    skdot/.style={circle,fill,inner sep=1.35pt},
    every label/.append style={font=\small, label distance=-2.5pt},
    skarrow/.style={
        postaction={decorate},
        decoration={markings, mark=at position #1 with {\arrow{Stealth[length=2.2mm]}}}},
    skarrow/.default=0.5,
]

\pgfmathsetmacro{\skpanelsep}{7.6}   
\pgfmathsetmacro{\sklen}{6.0}        
\pgfmathsetmacro{\skgap}{0.55}       
\pgfmathsetmacro{\skret}{2.2}        
\pgfmathsetmacro{\skvlen}{3.4}       

\pgfmathsetmacro{\skpA}{0.10}
\pgfmathsetmacro{\skpB}{0.32}
\pgfmathsetmacro{\skpC}{0.54}
\pgfmathsetmacro{\skpD}{0.76}

\begin{scope}[shift={(0,0)}]
  \draw[skarrow=0.43] (0,0) -- (\sklen,0);          
  \draw (\sklen,0) -- (\sklen,-\skgap);             
  \draw[skarrow=0.5] (\sklen,-\skgap) -- (0,-\skgap);

  \node[left] at (0,0)       {$\lvert\psi\rangle$};
  \node[left] at (0,-\skgap) {$\langle\psi\rvert$};

  \node[skdot,label={above:$W(t_4)$}] at ({\skpA*\sklen},0) {};
  \node[skdot,label={above:$W(t_3)$}] at ({\skpB*\sklen},0) {};
  \node[skdot,label={above:$V(t_2)$}] at ({\skpC*\sklen},0) {};
  \node[skdot,label={above:$V(t_1)$}] at ({\skpD*\sklen},0) {};
\end{scope}

\begin{scope}[shift={(\skpanelsep,0)}]
  \draw[skarrow=0.43] (0,0) -- (\sklen,0);
  \draw (\sklen,0) -- (\sklen,-\skgap);
  \draw[skarrow=0.5] (\sklen,-\skgap) -- (0,-\skgap);
  \draw (0,-\skgap) -- (0,{-\skgap-\skret});        
  \node[below] at (0,{-\skgap-\skret}) {$t_0 - i\beta$};

  \node[skdot,label={above:$W(t_4)$}] at ({\skpA*\sklen},0) {};
  \node[skdot,label={above:$W(t_3)$}] at ({\skpB*\sklen},0) {};
  \node[skdot,label={above:$V(t_2)$}] at ({\skpC*\sklen},0) {};
  \node[skdot,label={above:$V(t_1)$}] at ({\skpD*\sklen},0) {};
\end{scope}

\begin{scope}[shift={({2*\skpanelsep},0)}]
  \draw[skarrow=0.5] (0,0) -- (0,-\skvlen);
  \node[above] at (0,0)        {$t_0$};
  \node[below] at (0,-\skvlen) {$t_0 - i\beta$};

  \node[skdot,label={right:$W(t_4)$}] at (0,{-0.20*\skvlen}) {};
  \node[skdot,label={right:$W(t_3)$}] at (0,{-0.38*\skvlen}) {};
  \node[skdot,label={right:$V(t_2)$}] at (0,{-0.58*\skvlen}) {};
  \node[skdot,label={right:$V(t_1)$}] at (0,{-0.76*\skvlen}) {};
\end{scope}

\end{tikzpicture}%
}
\resizebox{\skscale\wd0}{!}{\usebox0}
\caption{Schwinger-Keldysh contours for
$\langle \psi| V(t_1)V(t_2) W(t_3)W(t_4)|\psi \rangle$.
\emph{Left:} the folded Lorentzian in-in contour.
\emph{Middle:} the state $|\psi\rangle$ replaced by a thermal density matrix.
\emph{Right:} thermal correlator continued to Euclidean time}
\label{fig:SK}
\end{figure}
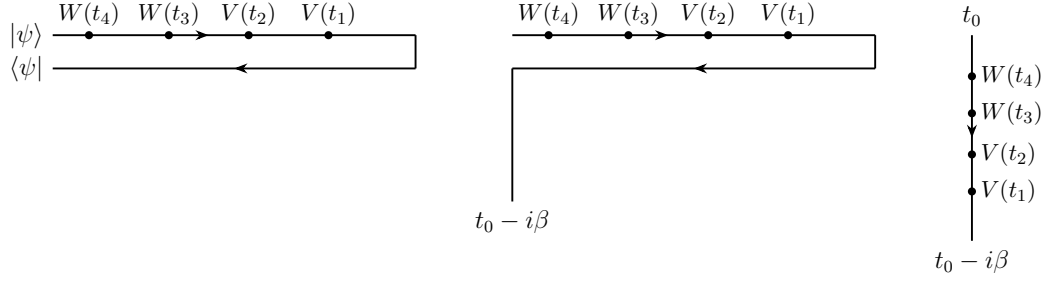

\subsection{Correlation functions of Hawking radiation: results}

Putting together the pieces presented above, we can now deduce results for the late-time correlation functions on $\Ic^+$ to first order in the effective coupling $g= G/(\pi r_0^3 T_H)$, corresponding to the diagrams in Figure  \rff{fig:vertices-pair}.

\subsubsection{Connected four-point function}

The connected thermal four-point function in AdS$_2$ is given by \rf{d30}.  Recall that we introduced two massless fields $V$ and $W$ to simplify the counting of diagrams.  Restoring the tildes and continuing to Lorentzian signature, this corresponds to
\eq{f19}{  \lim_{z\rt 0} \langle \p_z \Vt(z,\ttil_1) \p_z \Vt(z,\ttil_2) \p_z \Wt(z,\ttil_3) \p_z \Wt(z,\ttil_4)\rangle_c = {g\over 4\pi^3}{  \left( -2+ \ttil_{12} \coth {\ttil_{12} \over 2}  \right) \left( -2+ \ttil_{34} \coth {\ttil_{34} \over 2}  \right) \over  \big( 2 \sinh {\ttil_{12}\over 2}\big)^2  \big( 2 \sinh {\ttil_{34}\over 2}\big)^2 }~.}
Note that we are considering the time-ordered correlator,  $t_1>t_2>t_3>t_4$. 
On the right-hand side, we can replace $\ttil \rt \tilde{u}$ since these are equal at $z=0$.   Now we can apply the same logic as discussed below \rf{f11}. On the left hand side we can use  $\p_{\tilde{u}}\tilde{V}= {1\over 2}\p_z \tilde{V}$, and likewise for $\tilde{W}$.  Converting also the original untilded $u$ coordinate, we have 
%
\eq{f20}{ &\langle \p_u \Vt(u_1)\p_u \Vt(u_2) \p_u \Wt(u_3)\p_u \Wt(u_4)\rangle_c \cr
& \quad ={(2\pi T_H)^4\over 16} {g\over 4\pi^3} {  \Big( -2+ 2\pi T_H u_{12} \coth (\pi T_H u_{12})  \Big) \Big( -2+ 2\pi T_H u_{34} \coth (\pi T_H u_{34}) \Big) \over  \big( 2 \sinh( \pi T_H u_{12})\big)^2  \big( 2 \sinh (\pi T_H u_{34})\big)^2 }~.\cr }
For comparison, the disconnected correlator at order $g^0$ is 
\eq{f21}{ \langle \p_u \Vt(u_1)\p_u \Vt(u_2) \p_u \Wt(u_3)\p_u \Wt(u_4)\rangle_d & =   {\pi T_H^2 \over 4\sinh^2 (\pi T_H u_{12} ) }   {\pi T_H^2 \over 4\sinh^2 (\pi T_H u_{34}) } }
For $T_H u_{12} \sim T_H u_{34} \sim O(1)$ we see that the connected part has a relative factor of $g$ compared to the disconnected part, and hence becomes important near extremality.  On the other hand, if $T_H u_{12} \sim T_H u_{34}  \ll 1$ and $u_{12}\sim u_{34} \sim r_0$  then the connected part is suppressed by a factor of $G r_0 T_H^3$, which is small near extremality. So to see the large effects in the four-point function due to gravitational interactions in the near-horizon region, one must perform measurements over a length/time scale comparable to the thermal wavelength, which grows as we approach the extremal limit.   
 
Following MSY \cite{Maldacena:2016upp}, we can give a simple physical interpretation of the result \rf{f20}.   A black hole of mass $M$ has a nominal temperature $T_H$, but near extremality this notion loses its sharp meaning, and we should instead think of  $T_H$ as fluctuating.  The connected four-point function computed above can be thought of as resulting from correlated temperature fluctuations acting on the product of the leading-order two-point functions. More precisely, the following relation holds 
\eq{f22}{ & \langle \p_u \Vt(u_1)\p_u \Vt(u_2) \p_u \Wt(u_3)\p_u \Wt(u_4)\rangle_c \cr
& \quad = {1 \over  C_Q} \left(  T_H{\p \over \p T_H} \langle \p_u \Vt(u_1)\p_u \Vt(u_2) \rangle  \right) \left( T_H{\p \over \p T_H} \langle \p_u \Wt(u_3)\p_u \Wt(u_4) \rangle  \right) ~,}
where $C_Q  = {dM\over dT_H}\big|_Q \approx 4\pi^2 \sqrt{G}Q^3 T_H$ is the heat capacity at fixed charge, which controls the size of thermal fluctuations. Thus, the four-point function provides a rather direct physical measure of the near-horizon fluctuations.

Recall that the four-point function was computed ignoring the greybody factor.  These can be incorporated by Fourier transforming the correlators to frequency space,  multiplying each field insertion by a transmission coefficient ${\cal T}(\omega)$, and then transforming back to position space. The same factors appear both in the leading and subleading order correlators and so do not affect their relative sizes.   Note that the relation \rf{f22} will be modified due to the dependence of ${\cal T}(\omega)$ on the temperature, though this dependence is absent at the order we work here.   

\subsubsection{One-loop two-point function}

Starting from \rf{d38} and applying the same procedure as for the four-point function gives 
%
%
%
\eq{f23b}{& \langle \p_u \phit(u_1)\p_u \phit(u_2) \rangle_{1-{\rm loop}}  = {  T_H^2 g \over 8\sinh^2 x } \Bigg[ \frac{-2x^2-2 \pi  i x+1- \cosh(2x) +(i\pi+2x) \sinh (2x)}{4\sinh^2 x }\notag\\
& \quad\quad\quad\quad\quad\quad\quad\quad\quad\quad\quad\quad\quad\quad\quad\quad\quad\quad -\left(-1+{x \over \tanh x}\right)\left(-1+{x+ i \pi \over \tanh x  } \right) \Bigg]\,, }
 where we defined $ x = \pi T_H u_{12}$.   As with the four-point function,  for $x\ll 1$ and $u_1-u_2 \sim r_0$ this is small compared to the tree-level result \rf{f14}.  But for $x \sim 1$ it is of order $g$ compared to  \rf{f14} and thus becomes large as $T_H \rt  0$.
%
%

The contribution to the energy flux is
\eq{f24}{ \tilde{T}^{1-{\rm loop}}_{uu} = 2\pi \lim_{u_1\rt u_2}  \langle \p_u \phit(u_1)\p_u \phit(u_2) \rangle_{1-{\rm loop}}  = {\pi g \over 8}T_H^2~. }
Since this is of order $g$ times the tree-level result, it becomes significant near extremality as $g$ grows.

To interpret the result \rf{f24}, we note that we expect the relation between the black hole mass and temperature to get corrected at order $g$.  Start from the result \rf{f9a} for the flux at tree-level,
\eq{f26a}{  \tilde{T}^{\rm tree}_{uu} = {GE\over 12 r_0^3}~,}
where $E= M  - {Q\over \sqrt{G}}$ is the energy above extremality.  At lowest order in $g$ we have 
\eq{f27a}{ E = {2\pi^2 r_0^3 \over G} T_H^2= {2\pi \over g }T_H~.}
We now allow for an order $g^0$ quantum shift of the black hole mass at fixed $T_H$.   At one-loop order, we assume that $\tilde{T}_{uu}$ is still given by \rf{f26a} but in terms of the corrected mass, which gives the one-loop correction to the flux as 
\eq{f28a}{ \tilde{T}^{1-{\rm loop}}_{uu}= {G\delta E\over 12 r_0^3}~. }
Equating this to the computed result in \rf{f24} gives
\eq{f29a}{ \delta E = {3\over 2} {\pi r_0^3 \over G}g T_H^2 = {3\over 2}T_H~.}
We therefore interpret the one-loop correction to the flux as simply reflecting a one-loop shift by ${3\over 2}T_H$ of the black hole mass at fixed temperature.   The corrected formula for the black hole mass is therefore deduced as
\eq{f29b}{ M = {Q\over \sqrt{G} }  +2 \pi^2 \sqrt{G} Q^3 T_H^2+\frac{3}{2} T_H+\ldots .}

Using standard formulas, this mass shift translates into the following one-loop correction to the classical partition function,
\eq{f29}{ \ln Z(T_H) = \ln Z_{\rm cl}(T_H) + {3\over 2} \ln T_H + \ldots \,.}
The $ {3\over 2} \ln T_H$ correction precisely matches the one-loop correction deduced from the Schwarzian path integral \cite{Stanford:2017thb}.
In that computation, the factor of $3/2$ is tied to the three gauge modes associated with an SL(2,R) gauge symmetry.  By contrast, in our derivation there is no residual gauge symmetry, and we only computed a one-loop correlator of the scalar field.   The agreement between these two computations deserves further scrutiny.  In particular, the assumption stated below \rf{f27a} needs to be justified.

\section{Discussion}
\label{discuss}

In this work, we developed and applied a gauge-fixed Hamiltonian formulation of the 
s-wave sector to compute gravitationally induced correlations in Hawking radiation from
near-extremal \RN collapse. In this formulation, the gravitational constraints
encode the metric fluctuations in an action for the scalar field alone, while neglecting the
interaction terms smoothly recovers Hawking's original free-field treatment. The main
physical result is that the outgoing radiation develops parametrically enhanced correlations
near extremality, controlled by the coupling $g= G/( \pi r_0^3 T_H)$.
 For operator separations of order
the thermal time \(1/T_H\), the connected four-point function is larger than a generic
perturbative correction by the factor \(g\), and therefore becomes important as extremality
is approached. By contrast, for short separations it remains small. The effect  is thus an infrared correlation induced by large-scale fluctuations
of the near-horizon geometry.

These results have no direct bearing on the problem of information loss, nor do they provide a
description of the final strongly coupled regime reached arbitrarily close to extremality.
Rather, they identify the first controlled breakdown of the free Hawking state due to
low-energy quantum gravitational fluctuations in the near-horizon region.

We now compare this Hamiltonian formulation with the Schwarzian quantum-mechanics
approach, discuss several supporting computations and extensions presented in the appendices,
and outline possible directions for further work.

\subsection{Hamiltonian versus Schwarzian approaches}

It is useful to step back to compare and contrast the Hamiltonian and Schwarzian approaches to near-extremal black hole dynamics.    The Schwarzian approach isolates and quantizes the universal near-AdS$_2$ boundary mode. The Hamiltonian approach instead employs a gauge-fixed Lorentzian description of the s-wave sector and derives the near-horizon enhancement from the gravitational constraints.  There are relevant distinctions both in the way that the decomposition of the full spacetime into near-horizon and asymptotic regions is handled, and in the quantum description of the near-horizon region.

Regarding the first issue, studying wave equations on stationary black hole spacetimes using matched asymptotic expansions yields, for appropriate parameter regimes, a mathematically precise notion of ``near" and ``far" regions.  For near-extremal black holes, the near region contains an AdS$_2$ factor.   One can therefore aim to ``solve" the theory in the AdS$_2$ region and then match to the far region where the equations are relatively simple.\footnote{A separate step, not directly relevant here, is to posit a dual description of the near-horizon region in terms of some strongly coupled CFT or related theory.} At the level of the path integral this factorization of the spacetime into regions can be subtle; see \cite{Kolanowski:2024zrq,Bac:2026eqj,Despontin:2026xzg} for examples and discussion.   If one further argues that the AdS$_2$ region is described by JT gravity coupled to some light fields, and further takes this theory to be described by a boundary Schwarzian quantum mechanics coupled to boundary bilocal operators, then one arrives at a computationally powerful framework. In the Euclidean thermal setting, and under the assumptions discussed below, the near-horizon theory can be solved essentially exactly; e.g., boundary correlators can be computed to all orders in the coupling \cite{Mertens:2017mtv,Yang:2018gdb,Iliesiu:2019xuh}.   

The Schwarzian formulation is less well adapted to collapse problems. To use it after black hole formation, one must supplement the near-horizon theory with a prescription for the state, and that state depends on the prior Lorentzian history.
By contrast, in the Hamiltonian approach there is one uniform description valid on the full spacetime,\footnote{Putting aside issues regarding the validity of the s-wave reduction.} which holds before the geometry has settled down to a stationary state.  The near-horizon region is distinguished by the fact that the effective coupling in this region can grow large near extremality, and so it can make sense to only consider interactions that occur in this region. But in principle nothing requires this split to be made, and computations in the full Lorentzian spacetime can be carried out using ordinary perturbation theory.  However, it has to be said that explicit computations of correlation functions are more involved in this framework; this is so even for tree-level correlators in thermal AdS$_2$, where we showed full agreement between the two descriptions.   

Next we turn to the quantum treatment of the near-horizon AdS$_2$ region in the two descriptions, starting with the Schwarzian.    The cleanest setting is for Euclidean (thermal) AdS$_2$.  To start, one argues that the near-horizon region is governed by JT gravity, where the action contains the piece $I_{\rm JT}  = -{1\over 16\pi G_2} \int\! d^2x \sqrt{g}\Phi ({\cal R}_2 +2)$, $\Phi$ being the dilaton.  The key step in the derivation of the Schwarzian theory is to view $\Phi$ as a Lagrange multiplier that imposes the constraint ${\cal R}_2=-2$ inside the path integral, restricting the metric to be AdS$_2$. The issue here is that,  from its higher-dimensional origin, the dilaton is related to the size of a sphere factor on which the theory is reduced, and as such is both real and bounded below.  However,  the Lagrange multiplier interpretation requires $\Phi$ to be integrated over the full imaginary axis; this choice of contour is typically regarded as being part of the definition of quantum JT gravity, rather than being derived from the higher-dimensional starting point.   Having granted this, the power of the Schwarzian approach relies on the fact that the integration space is the symplectic manifold $\operatorname{Diff}(S^1)/SL(2,\mathbb R)$ and is therefore equipped with a natural symplectic measure; see \cite{Stanford:2017thb} for a discussion of the measure and its UV divergences.   Bilocal operators in this theory, used to compute boundary correlators, similarly have a natural definition, facilitating their efficient evaluation \cite{Mertens:2017mtv,Yang:2018gdb,Iliesiu:2019xuh}.  In particular, under these rules correlation functions appear to be well-defined order-by-order in perturbation theory \cite{Griguolo:2021zsn}. We simply remark that these features, while natural and powerful, are not derived from a higher-dimensional starting point, nor fixed by imposing a symmetry,   but rather taken as part of the definition of the reduced theory.  

In contrast, in the Hamiltonian approach, perturbative evaluation of loop-level boundary correlators in Euclidean AdS$_2$ encounters UV divergences; we saw a mild example of this for the two-point function at one-loop, but more generally we expect that our Hamiltonian theory is not UV complete. This is hardly surprising given that the starting point of four-dimensional Einstein gravity shares this property, and the reduction to the s-wave also discards degrees of freedom which are important at short distances.   Relatedly, a key difference is that in the Hamiltonian formulation there is no analog of a Lagrange multiplier constraint fixing the metric to be precisely AdS$_2$; the restriction to AdS$_2$ requires the equations of motion and not just the gravitational constraints, and so does not hold inside the path integral.

This raises two related questions. First, can one choose a measure and counterterms in the Hamiltonian path integral such that loop-level Hamiltonian correlators agree with Schwarzian correlators? Second, which aspects of near-horizon physics are universal consequences of the AdS$_2$ throat, and which depend on the UV completion or on the precise higher-dimensional embedding? We leave these questions for future work.

\subsection{Supporting computations and extensions in the appendices}

The appendices contain both technical material used in the main text and a few extensions of
the Hamiltonian method. Appendices \rff{corr}-\rff{grey} collect some technical details: Appendix \rff{corr} gives the Fourier-space manipulations needed for the
dressed bilocal correlators; Appendix \rff{variational} reviews the variational principle and boundary
terms for massive fields in the reduced Hamiltonian description; and Appendix \rff{grey} reviews the derivation of transmission and reflection coefficients at low frequency for the near-extremal \RN solution.   Appendices \rff{otoc}-\rff{ads3} illustrate how the Hamiltonian method can be applied to related near-horizon phenomena.

In Appendix \rff{otoc} we give a Lorentzian derivation of the leading exponential growth of an out-of-time-order correlator.   Although the result is standard,  it is still illuminating to see it play out in our Hamiltonian approach.   We work this out directly in Lorentzian signature, rather than by analytically continuing the appropriate Euclidean correlator.  As expected, the exponential Lyapunov growth arises from high-energy scattering near the horizon.

In Appendix \rff{charged}  we consider a charged scalar field in AdS$_2$ and work out the corresponding near-extremal enhanced interaction, now mediated by a ``phase mode" rather than the Schwarzian, which is again a well-known effect.  The main novelty here is that the enhancement mechanism is a bit different than in the gravitational case, and arises from the need to impose a cutoff on a divergent radial integral.

Finally, in Appendix \rff{ads3} we work out the s-wave reduction of AdS$_3$ gravity coupled to a scalar field and derive the corresponding Hamiltonian description and its near-horizon limit.  This is technically simpler than the reduction of four-dimensional Einstein-Maxwell theory.  The result in the near-horizon limit is the same, where the dimensionless coupling now takes the form $g={G_3\over \pi \ell^2 T_H}$,  $\ell$ being the AdS$_3$ radius. This example should mainly be viewed as a useful check
on the formalism: unlike the \RN case, the low-temperature BTZ limit has
$r_h \rt 0$, so the s-wave reduction is not parametrically controlled.

 \subsection{Future directions}

We close with a few comments on possible extensions of this work that we hope to examine in the future.   In this work, we restricted attention to the simplest observables, namely late-time correlators.  The simplicity of these is tied to their universality:  their form is independent of the details of the collapse process.   It would, however, be worthwhile, and should be feasible, to extend these computations to earlier times, as it would be interesting to see how the strong coupling turns on as the near-horizon region develops. Similarly, rather than starting with the vacuum state at $\Ic^-$, we could send in some particles and study their effects on the outgoing radiation, presumably finding the same sort of chaotic response found in \cite{Polchinski:2015cea}.   

Another avenue to consider, relevant for generic black holes, is to compute the high-energy tail of the Hawking spectrum, where the energy of a single quantum is large enough to have significant backreaction on the geometry.   A model for this process, also using Hamiltonian methods but in a different context, was developed in \cite{Kraus:1994fj,Keski-Vakkuri:1996wom}.  
The main result of those works is that in the emission probability the usual Boltzmann factor $e^{-\beta \omega}$ is replaced by a factor of $e^{-\Delta S}$,  where $\Delta S= S(M)-S(M-\omega)$ is the change in the black hole entropy.    Since the deviation from the Boltzmann factor is due to the gravitational self-interaction of the emitted quantum, it might be possible to verify this using the methods developed here.      

In another direction, while we focused here on (Minkowski space and AdS$_2$) boundary correlators, it would be interesting to extend this to bulk correlators.  Since we have fully fixed the gauge, such correlators are physical; they can equivalently be thought of as ``dressed observables" in the sense discussed in \cite{Cheung:2026euf}.  For example, we can use their singularity structure to study the extent to which lightcones become fuzzy due to gravitational fluctuations.  See \cite{Blommaert:2019hjr,Sivaramakrishnan:2025srr,Franken:2026bff,Freivogel:2026ofo,Freivogel:2026bsx} for related work.    To this end, it would be useful to work out the scalar action in an alternative gauge choice in which spatial slices cross the event horizon.  This would additionally allow for computation of two-sided correlators in eternal black hole spacetimes.     Finally,  to sharpen our understanding of the regime of validity of the s-wave reduction, it would be worthwhile to incorporate small fluctuations that depart from spherical symmetry and to work out their contributions to correlation functions.

\section*{Acknowledgments}

We would like to thank Alex Belin and  Ben Freivogel for helpful discussions. The work of SB and SK is supported by the Mani L. Bhaumik Institute for Theoretical Physics.

\appendix

\section{Technical details for correlator computations}
\label{corr} 

In this appendix, we start from the Schwarzian action coupled to dressed bilocal operators and use this to write the corresponding tree-level four-point correlator and one-loop two-point correlator as Fourier sums, allowing for comparison with the corresponding Hamiltonian expressions for these correlators. 

\subsection{Schwarzian dressed bilocal}

The action for a free massive scalar field  on a fixed Euclidean AdS$_2$ background, with boundary condition $\chi(\tau)$ is given by
\eq{j1}{ I_\phi =-D_\Delta  \int\! d\tau_1 d\tau_2 {\chi(\tau_1)\chi(\tau_2) \over \left( 2\sin {\tau_{12}\over 2}\right)^{2\Delta} }~,\quad \tau_{12} = \tau_1 -\tau_2 \,,}
with 
\eq{j2}{ D_\Delta =\frac{\left(\Delta-\frac{1}{2}\right) \Gamma(\Delta)}{\sqrt{\pi} \Gamma\left(\Delta-\frac{1}{2}\right)}~,\quad  D_1 = {1\over 2\pi}~.}
The Schwarzian dressed version defines the bilocal operator
\eq{j3}{ I_{\rm bilocal} = -D_\Delta  \int\! d\tau_1 d\tau_2  \left(  {T'  (\tau_1) T'(\tau_2) \over  \left( 2\sin  {T(\tau_1)-T(\tau_2) \over 2}\right)^{2} }\right)^\Delta \chi(\tau_1)\chi(\tau_2)  \,, }
with 
\eq{j4}{ T(\tau) = \tau+\eps(\tau)~.}
We now set $\Delta =1$  and take $\chi(\tau) = \delta(\tau-\tau_1)+\delta(\tau-\tau_2)$, keeping just the cross terms in the expansion of $ \chi(\tau_1)\chi(\tau_2) $,
\eq{j4a}{ I_{\rm bilocal}[\tau_1,\tau_2]  =  - {1\over \pi}    {T'  (\tau_1) T'(\tau_2) \over  \left( 2\sin  {T(\tau_1)-T(\tau_2) \over 2}\right)^{2} }~.}

Perturbation theory is carried out by expanding in $\eps(\tau)$.  

\subsection{First order expansion} 

At first order we have 
\eq{j5}{
I^{(1)}_{\rm bilocal}[\tau_1,\tau_2]
=- {G_0(\tau_{12})\over \pi} 
\left[
\epsilon^{\prime}\left(\tau_1\right)
+\epsilon^{\prime}\left(\tau_2\right)
-\left(\epsilon\left(\tau_1\right)-\epsilon\left(\tau_2\right)\right)
\cot { \tau_{12} \over 2}
\right]
}
with
\eq{j6}{ G_0(\tau_{12})  =   {1\over \left( 2 \sin {\tau_{12} \over 2} \right)^2}~.}
For our purposes, we need to write \rf{j5} as a Fourier sum.  Using $ G_0^{\prime}(\tau)=-\cot \left(\frac{\tau}{2}\right)  G_0(\tau) $  we first write 
\eq{j7}{    I_{\mathrm{bilocal}}^{(1)}[\tau_1,\tau_2]=-{G_0(\tau_{12}) \over \pi} \left(\epsilon_1^{\prime}+\epsilon_2^{\prime}\right)+{1\over \pi} \left(\epsilon_1-\epsilon_2\right) G_0^{\prime}(\tau_{12})~,     }
Inserting the Fourier expansions  (the regulator $\delta$ is taken to zero at the end) 
\eq{j8}{ \epsilon(\tau)=\sum_k \epsilon_k e^{-i k \tau}, \quad G_0(\tau)=-\sum_{n=1}^{\infty} n e^{-n\delta}  e^{-i n \tau} }
we obtain
\eq{j9}{    I_{\mathrm{bilocal}}^{(1)}[\tau_1,\tau_2] =-   {i\over \pi}    \sum_{|k|>1}  e^{i k \tau_2} \eps_{-k}    \sum_{n=1}^\infty    \Big( \left[n(n-k) e^{i(k-n) \tau_{12}}-n(n+k) e^{-i n \tau_{12}}\right] \Big) ~.          }
Splitting the   $k$-sum into two parts corresponding to the sign of $k$ and doing some algebra, we end up with 
\eq{j10}{    I_{\mathrm{bilocal}}^{(1)}[\tau_1,\tau_2] =- { i\over \pi}  \sum'_{n_1,n_2} n_1 n_2 e^{in_1 \tau_1 +in_2 \tau_2} \eps_{-(n_1+n_2)}\,, }
where $\sum'$ means sum over all $n_{1,2}$ with $n_1 n_2>0$.     This is the result needed to verify \rf{d29}.

\subsection{Second order expansion} 

For the one-loop two-point function, we need the expansion of the bilocal  \rf{j4a} to second order. It is convenient to first use 
\eq{j11}{ 
 I_{\rm bilocal}[\tau_1,\tau_2] =- {1\over \pi}\partial_{\tau_1} \partial_{\tau_2} \ln \left[2 \sin \left(\frac{T\left(\tau_1\right)-T\left(\tau_2\right)}{2} \right)\right] ~.}
We then expand the logarithm to second order,
\eq{j12}{ \ln \left[2 \sin \frac{T\left(\tau_1\right)-T\left(\tau_2\right)}{2}\right]  & =  O(\eps^0) + O(\eps) - {1\over 2} G_0(\tau_{12} ) \big(\eps_1 -\eps_2\big)^2 + \ldots }
%
To compute the two-point function, we need the expectation value of $\big(\eps_1 -\eps_2\big)^2$ computed with respect to the Schwarzian action $I_{\rm Sch} = {2\pi \over g} \sum_{|k|>1} k^2 (k^2-1)\eps_k \eps_{-k}$.  This gives
\eq{j13}{ \langle \eps(\tau_1)\eps(\tau_2)\rangle = {g\over 4\pi} \sum_{|k|>1} {e^{ik \tau_{12}} \over k^2 (k^2-1)}~.}
When we evaluate $\langle \big(\eps_1 -\eps_2\big)^2\rangle$ we have both self and non-self contractions, and it is useful to separate these:
\eq{j14}{ \langle \big(\eps_1 -\eps_2\big)^2\rangle & = \langle \big(\eps_1 -\eps_2\big)^2\rangle_{\rm self}  +  \langle \big(\eps_1 -\eps_2\big)^2\rangle_{\rm non-self} \cr
& =  {g\over 2\pi} \sum_{|k|>1} {1 \over k^2 (k^2-1)}  - {g\over 2\pi}  \sum_{|k|>1} {e^{ik \tau_{12}} \over k^2 (k^2-1)}~.}
Using \rf{j8} we then compute $\langle I^{(2)}_{\rm bilocal}[\tau_1,\tau_2]\rangle $, distinguishing the two contributions,
\eq{j15}{ \langle I^{(2)}_{\rm bilocal}[\tau_1,\tau_2]\rangle_{\rm self} & =- {g\over 8\pi^2} \sum_{n, |k|>1}    { |n|^3 \over k^2 (k^2-1)} e^{-in\tau_{12} }    \cr
\langle I^{(2)}_{\rm bilocal}[\tau_1,\tau_2]\rangle_{\rm non-self} & =   {g\over 8\pi^2} \sum_{n, |k|>1}  { |n| (k-n)^2 \over k^2(k^2-1)  } e^{i(k-n)\tau_{12}  }~.
 }
After some algebra, we find that the sum of these contributions can be written in the form
\eq{j16}{ \langle I^{(2)}_{\rm bilocal}[\tau_1,\tau_2]\rangle & = {g\over 4\pi^2} \sum_{m=1}^{\infty} \sum_{r=1}^{\infty} \frac{r m^2}{(m+r)^2\left[(m+r)^2-1\right]}\left(e^{i m \tau_{12}}+e^{-i m \tau_{12}}\right)~.}
This is the two-point function at one-loop, and we see that it agrees with our result \rf{d33}.

\section{Variational principle and boundary terms for massive fields}
\label{variational}

Before adding a counterterm, the reduced action for a massive scalar field in the near-horizon region is 
\eq{h1}{  S = \int\! d{t} \int_1^\infty d{r}  \Big[  \pi_\phi \dot{\phi} - \Big( {f}({r}) {h}({r}) +{V}({r})\Big) e^{-g \int_{{r}}^\infty d{r}' {h}({r}')} \Big]   }
with 
\eq{rv5}{ {h} = {1\over 2}( \pi_\phi^2 + \phi'{}^2)~,\quad {f}  = {r}^2-1~,\quad V = {1\over 2} \Delta(\Delta-1) \phi^2~.}
The corresponding Hamiltonian is 
\eq{h2}{ H & = H_0 + H_{\rm ct} \cr
&  =\int_1^\infty\! dr \Big( {f}({r}) {h}({r}) +  {V}({r})\Big) E(r)    + H_{\rm ct}~,}
with 
\eq{h3}{ E(r) = e^{-g \int_{{r}}^\infty d{r}' {h}({r}')}~. }
Varying the Hamiltonian gives a bulk term, which contributes to the equations of motion and is omitted here, along with a boundary term
\eq{h4}{ \delta H = \lim_{r\rt \infty}  \Big( F(r) \phi'(r)   \delta \phi(r) \Big)  + \delta H_{\rm ct}}
with 
\eq{h5}{ F(r) = f(r) E(r) -g \int_1^r \! dr' \Big( f(r') h(r') + V(r')\Big) E(r')~. }
The on-shell variation of the action is $\delta S = -\int\! dt\delta H$. For the solutions of interest we have asymptotics
\eq{h6}{ F(r) & \approx   r^2 +O(r^0) \cr
  \phi(r,t) &\approx r^{\Delta-1}\chi(t) + r^{-\Delta} \phi_\Delta(t) ~.}
Using this, we evaluate
\eq{h7}{ \delta H =  (\Delta-1)^2 r^{2\Delta-1}  \chi \delta \chi + (\Delta-1) \chi \delta \phi_\Delta -\Delta \phi_\Delta \delta \chi  + \delta H_{\rm ct}~.}
For a good variational principle in which we fix $\chi$, there should be no $\delta \phi_\Delta$ term.  We therefore take 
\eq{h8}{ H_{ct} = - (\Delta-1) \chi \phi_\Delta \,,}
 giving 
\eq{h9}{ \delta H   = -(2\Delta-1) \phi_\Delta \delta \chi ~.           }
The on-shell variation of the action is then 
\eq{h10}{ \delta S = \int\! dt \Oc_\Delta \delta \chi \,,}
where the operator conjugate to $\chi$ is thereby identified as 
\eq{h11}{  \Oc_\Delta = (2\Delta-1)\phi_\Delta\,,}
which is a standard formula.  Note that $H_{\rm ct}$ vanishes for a massless scalar ($\Delta=1$), so it plays no role in most of the computations in this paper. 

\section{Transmission and reflection coefficients}
\label{grey}

We sketch the derivation of some relevant aspects of solutions of the scalar wave equation on the near-extremal black hole background, with the main goal being to explain the emergence of an approximate reflecting boundary condition at the boundary of the near-horizon AdS$_2$ region.  Standard references for this sort of analysis include \cite{Page:1976df,Unruh:1976fm,Das:1996we,Maldacena:1996ix,Cvetic:1997uw,Cvetic:1997xv}.

We consider the wave equation \rf{f2} 
\eq{q1}{   \big(-\p_t^2 + \p_{r_*}^2 -V(r_*)\big) \phit=0~,\quad  V(r_*) = { (r-r_+)(r-r_-)[r_+(r-r_-)+r_-(r-r_+)]\over r^6}~.}
Assuming time-dependence $e^{-i\omega t}$, we are interested in solutions of the form 
\eq{q2}{ \phit ~\approx~   \left\{ \begin{array}{cc}
  e^{i\omega r_*} + {\cal R}(\omega) e^{-i\omega r_*}~, & r_* \rt -\infty  \\
 {\cal  T}(\omega) e^{i\omega r_*}~,  &  r_* \rt \infty 
\end{array} \right. \,.}
The regime of interest is $\omega r_0,  T_H r_0 \ll 1$.    We also note that the coordinates $(z,\ttil) $ used in the AdS$_2$ region ($r-r_+ \ll r_+$) are related to $(r_*,t)$ as 
\eq{q3}{ r_* \approx -{z\over 2\pi T_H} + {\rm const} ~,\quad  t = {\ttil\over 2\pi T_H}~.}
It will be convenient to use $z$ in the AdS$_2$ region, where the solution then takes the form 
\eq{q4}{  \phit = e^{-i \omt z } + {\cal R}(\omega) e^{i\omt z}~,\quad \omt = {\omega \over 2\pi T_H}\,. }
Here we redefined  ${\cal R}$ by a convention-dependent phase given by the constant in \rf{q3}.
For small $z$ we therefore have 
\eq{q5}{ \phit \approx (1+{\cal R}) - i\omt (1-{\cal R}) z +O(z^2)~,\quad  \omt z \ll 1~. }
On the other hand, in the far region $r\gg r_0$ we have
\eq{q6}{ \phit \approx  {\cal T}(1+i\omega r+ \ldots)~,\quad \omega r \ll 1~.}
The near and far regions both overlap with an intermediate matching region where the solution takes the form
\eq{q7}{\phit \approx  Ar - B - {Br_0 \over r-r_0}}
obtained by dropping the $\omega$ dependence in the wave equation, and is valid in the region $r_0^2 T_H \ll r-r_0 \ll {1\over \omega}$.  Matching to \rf{q6} gives $A=i\omega {\cal T}$ and $B=-{\cal T}$.
To match to \rf{q5} we use that at small $z$ we have $r-r_0 = \lambda\rho_h \coth z \approx {2\pi r_0^2 T_H \over z}$ so that \rf{q7} can be written $\phit = Ar_0-B  -{B\over 2\pi r_0 T_H}z $, yielding $Ar_0-B = 1+{\cal R}$ and  $B = i\omega r_0 (1-{\cal R})$. Solving gives
\eq{q8}{ {\cal R} = -1-2i \omega r_0+ 4(\omega r_0)^2 + \ldots~,\quad {\cal T}  = -2i \omega r_0 + \ldots~,}
which obey $|{\cal R}|^2 + |{\cal T}|^2=1$ to the order considered. 

From \rf{q5} the behavior near the AdS$_2$ boundary is therefore 
\eq{q9}{ \phit \approx  \phi_0  + \phi_1 z + O(z^2) }
with 
\eq{q10}{ {\phi_0 \over \phi_1 } \approx 2\pi r_0 T_H \ll 1~.}
The smallness of $\phi_0$ compared to $\phi_1$ near-extremality implies an approximate Dirichlet (or reflecting) boundary condition $\phit|_{z=0} \approx 0$.  This is the same boundary condition assumed in our computation of AdS$_2$ boundary correlators, and justifies our use of these correlators. 
  
\section{OTOC and Lyapunov exponent}
\label{otoc}

Out-of-time-order correlators serve as useful diagnostics of chaos, e.g., \cite{Maldacena:2015waa}.  The exponential Lyapunov growth in time of such correlators arises in gravity from high-energy collisions near a black hole horizon \cite{Shenker:2013pqa,Shenker:2013yza,Shenker:2014cwa}.  Following MSY \cite{Maldacena:2016upp}, we consider a Lorentzian finite temperature AdS$_2$   correlator of the form 

\eq{w1}{ G_4= \left\langle V(t_3) W(t_1) V(t_4) W({t_2})\right\rangle_\beta~, }
with 
\eq{w1a}{ t_1=b+\that~,\quad t_2=\that~,\quad t_3= a~,\quad t_4= 0~,\quad a,b\sim \beta~,\quad \that\gg \beta~.}
 At first order in $G$ the correlator exhibits the Lyapunov growth
\eq{w2}{  G_4(\that) \sim e^{\frac{2\pi}{ \beta}\that },\quad \that \gg \beta~. }
We henceforth set $\beta =2\pi$.

The result \rf{w2} can be derived by analytic continuation of the corresponding Euclidean correlator, but it is more physically illuminating to work directly in Lorentzian signature in order to expose the scattering interpretation.  This was done in \cite{Shenker:2013pqa,Shenker:2013yza,Shenker:2014cwa}; here we reproduce this in our Hamiltonian approach.

\subsection{Setup}

Our Hamiltonian has the quartic interaction vertex 
\eq{w3}{ H_I = -g\int\! dt \int_0^\infty \! dz \int_0^{z} \! dz' \sinh^2 z' \Big[  T^V_{tt}(t,z)T^W_{tt}(t,z')+T^W_{tt}(t,z)T^V_{tt}(t,z')\Big]\,,}
which we use to compute the boundary four-point function.   Here $T_{tt}^V = {1\over 2}( \dot{V}^2+V'{}^2)$, and likewise for $T_{tt}^W$.  We are working in the metric 
\eq{w4}{ ds^2 = {-dt^2 +dz^2 \over \sinh^2 z}~.}
The exponential growth \rf{w2} will arise from the growth of the $\sinh^2 z'$ factor near the horizon, $z' \rt \infty$.

First-order time-dependent perturbation theory yields the expression,
\eq{w5}{ G_4 & =  -i \int_{t_2}^{t_4} \! dt  \langle  V_I(t_3)  W_I(t_1) V_I(t_4) H_I(t)\rangle W_I(t_2)_\beta     \cr
&  -i \int_{t_4}^{t_1} \! dt  \langle  V_I(t_3)  W_I(t_1)  H_I(t) V_I(t_4) W_I(t_2)\rangle_\beta  \cr 
 & -i \int_{t_1}^{t_3} \! dt  \langle  V_I(t_3) H_I(t) W_I(t_1)   V_I(t_4) W_I(t_2)\rangle_\beta  \cr  
 &  -i \int_{t_3}^{t_4} \! dt  \langle   H_I(t)  V_I(t_3) W_I(t_1)   V_I(t_4) W_I(t_2)\rangle_\beta  \cr   
  & -i \int_{t_4}^{t_4-i\beta } \! dt  \langle   H_I(t)  V_I(t_3) W_I(t_1)   V_I(t_4) W_I(t_2)\rangle_\beta \,, }
where $I$ stands for interaction picture, which is to say that the fields are all free. 
This is illustrated in Figure \rff{SK2}.
%
%


\begin{figure}[H]
\centering
\begin{tikzpicture}[
    x=1cm,y=1cm,
    line cap=round,line join=round,
    every node/.style={font=\small}
]

\coordinate (A) at (0,3.00);    
\coordinate (B) at (5.00,3.00); 
\coordinate (C) at (5.00,2.55); 
\coordinate (D) at (0,2.55);    
\coordinate (E) at (0,2.10);    
\coordinate (F) at (6.60,2.10); 
\coordinate (G) at (6.60,1.65); 
\coordinate (H) at (0,1.65);    
\coordinate (I) at (0,0.25);    

\draw (A) -- (B);
\draw (B) -- (C);
\draw (C) -- (D);
\draw (D) -- (E);
\draw[blue] (E) -- (F);
\draw (F) -- (G);
\draw (G) -- (H);
\draw[->] (H) -- (I);

\draw[->] (2.20,3.00) -- (2.80,3.00);
\draw[->] (2.80,2.55) -- (2.20,2.55);
\draw[->,blue] (3.00,2.10) -- (3.60,2.10);
\draw[->] (3.60,1.65) -- (3.00,1.65);

\fill (D) circle (1.5pt);
\node[left=4pt] at (D) {$V(t_4)$};

\fill (B) circle (1.5pt);
\node[right=4pt] at (B) {$W(t_2)$};

\fill (F) circle (1.5pt);
\node[right=4pt] at (F) {$W(t_1)$};

\fill (2.45,1.65) circle (1.5pt);
\node[below=2pt] at (2.45,1.65) {$V(t_3)$};

\node[left=2pt] at (I) {$t_4-i\beta$};

\end{tikzpicture}
\caption{Schwinger--Keldysh contour for the OTOC. The interaction vertex is integrated over the contour according to \rf{w5}.  The blue segment gives the exponentially growing contribution.}
\label{SK2}
\end{figure}
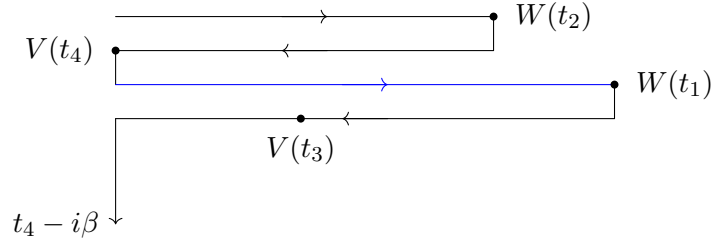

The fields in the interaction vertex are Wick contracted with the external fields using the bulk-boundary propagator. We define two distinct propagators, distinguished by whether the bulk time $t$ or boundary time $t_i$ is further along the Schwinger-Keldysh contour, as this determines the operator ordering.    We have
\eq{w6}{  K^\pm\left(t,z;t_i\right)& =- {1\over 4\pi} {\sinh z \over \sinh \left({x^+-t_i\mp i\eps \over 2}\right) \sinh \left({x^--t_i\mp i\eps \over 2}\right)  } \,, }
where 
\eq{w7}{ x^\pm = t\pm z~.}
We use $K^+ /K^-$ if $t/t_i$ is further along the contour.

As will become clear, the growing term comes from Wick contractions involving two $K^+$'s and two $K^-$'s, which occurs when the interaction vertex lies on the blue segment in Fig. \rff{SK2}. The contribution to the four-point function, which we denote by $g_4(\that)$, is 
\eq{w8}{ g_4(\that) & =  -ig \int_0^{b+\that} \! dt  \int_0^\infty \! dz \int_0^{z} \! dz' \sinh^2 z' \Big[  T^V_{tt}(t,z)T^W_{tt}(t,z')+T^W_{tt}(t,z)T^V_{tt}(t,z')\Big]\,,}
with
\eq{w9}{ T^W_{tt}(t,z) &= {1\over 2} \Big(\p_t K^+(t,z;\that)\p_t K^-(t,z;b+\that) + \p_z K^+(t,z;\that)\p_z K^-(t,z;b+\that)  \Big) \cr
T^V_{tt}(t,z) &= {1\over 2} \Big(\p_t K^+(t,z;0)\p_t K^-(t,z;a) + \p_z K^+(t,z;0)\p_z K^-(t,z;a)  \Big)\,. }

\subsection{Computation }

The $e^{\that}$ growth comes from high-energy scattering near the horizon, where it is amplified by the blue shift.  So the relevant region that contributes to the integral in \rf{w8} is where $z,z' \gg 1$ and we are within $O(1)$ of the lightcone singularities of the propagators.  So we first of all use 
\eq{w10}{ \sinh^2 z' \approx {1\over 4}e^{2z'}~. }
Next we take the lightcone approximations of \rf{w6}. For the past lightcone emanating from $\that$, which is $x^+ = t+z =\that$, we have $x^- -\that = t-z-\that =-2z$, and therefore  
\eq{w11}{  K^\pm\left(t,z;\that\right)& \approx {1\over 4\pi} {1 \over \sinh \left({x^+-\that\mp i\eps \over 2}\right)  }  \approx {1\over 2\pi}  {1\over x^+-\that\mp i\eps}~. }
Similarly, for the lightcone emanating from $0$, which is $x^- = t-z=0$, we have $x^+= t+z = 2z$ so that 
\eq{w12}{  K^\pm\left(t,z;0\right)& \approx -{1\over 4\pi} {1 \over \sinh \left({x^-\mp i\eps \over 2}\right)  }  \approx -{1\over 2\pi}  {1\over x^-\mp i\eps}~. }
These formulas give 
\eq{w13}{ T^W_{tt}(t,z) &\approx  {1\over 4\pi^2}   {1\over (x^+-\that -i\eps)^2 (x^+-\that-b +i\eps)^2} \cr
T^V_{tt}(t,z) &\approx  {1\over 4\pi^2}   {1\over (x^- -i\eps)^2 (x^--a +i\eps)^2}~. }
Now we have to be careful to restrict the integration domain to where the simplifications above are valid. We start from the stated assumptions
\eq{w14}{  z \sim z' \gg 1~,\quad    t+z-\hat{t}= O(1)~,\quad t-z = O(1)~. }
These imply  
\eq{w15}{ t\sim z \sim z' \sim {\that\over 2} +O(1)~.}
We will therefore take 
\eq{w16}{ \sinh^2 z'  \approx {1\over 4}e^{\that} }
and this is in fact the origin of the $e^{\that}$ growth. 

Now consider the $dz$ and $dz'$ integrals.   Given our comments above, these take the form 
\eq{w17}{ \int_{-L}^L {dz \over  (z -i\eps)^2(z - c +i\eps)^2 }}
where $c$ is an   $O(1)$  constant, and the endpoints $\pm L$ demarcate  the width of the light cone region.   Here is where we see the importance of the mixed $\pm i\eps$ terms.  If both $i\eps$'s come with the same sign, then the integral goes to zero for large $L$ (since we can close the contour), while in the mixed sign case it goes to a finite constant $\sim 1/c^3$.    Note that we think of $L$ as being a fixed, numerically large quantity, not scaling with $\that$.  Under this assumption, the $z$ and $z'$ integrals evaluate to constants, leaving us with the $t$ integral, which gives the $O(1)$ width  $L$.  

Putting these facts together,  we arrive at  
\eq{w18}{ g_4(\that)  \approx g e^{\that}~,\quad \that \gg 1  \,,}
where the numerical prefactor is not determined at the level of our analysis. This gives the expected Lyapunov exponent.

\section{Charged scalar and phase mode theory}
\label{charged}

The Hamiltonian approach can also be used to derive the phase mode theory describing the enhanced near-horizon interactions of a charged scalar field.   Here we sketch the derivation.  See \cite{Davison:2016ngz,Bulycheva:2017uqj,Gaikwad:2018dfc,Moitra:2018jqs,Sachdev:2019bjn} for prior related work, both in gravity and in the SYK model. To simplify, we take the scalar field to be charged under a different $U(1)$ than the $U(1)$ under which the \RN black hole is charged.    So we take the action 
\eq{g1}{ S = - {1\over 4\pi}\int\! d^4x \sqrt{-g} \left[  {1\over 4G} {\cal R} +{1\over 4} F_{\mu\nu}^2 +{1\over 4} f_{\mu\nu}^2  + |D \phi|^2\right]~,}
with $D_\mu  = \p_\mu + iqa_\mu$. Choosing the gauge $a_r=0$, the Hamiltonian works out to be 
\eq{g2}{ H_{ADM} =  M + \int_{r_h}^\infty\! dr' \left(  f(r') h(r')  +{\Qc^2 \over 2r'{}^2}  \right) e^{-\int_{r'}^\infty \! dr'' {2Gh(r'')\over r''}}~.  } 
Here $\Qc= \pi_{a_r}$ obeys the Gauss law constraint 
\eq{g3}{ \Qc'=   iq( \pi_\phi \phi - \pi_{\phi^*} \phi^*) \,, }
and we impose the boundary condition  $ \Qc(r_+) =0$ so that 
\eq{g4}{  \Qc(r)= \int_{r_+}^r \! dr'   iq( \pi_\phi \phi - \pi_{\phi^*} \phi^*)  ~. }
In order to focus on the electromagnetic effects, we now set $G=0$ in the exponential term of \rf{g2}.   

At lowest order in $q$, which is all we need for computing the tree-level four-point function, we have $\pi_\phi = \dot{\phi}^*$.   Taking the near-horizon limit, we obtain the action 
\eq{g5}{ S = \int \! dt \int_0^\infty dz \left( - |\p \phi|^2 - {1\over \sinh^2 z} \int_z^\infty \! dz' J(z') \int_z^\infty \! dz'' J(z'') \right) \,,}
with 
\eq{g6}{ J = iq ( \dot{\phi}^* \phi - \dot{\phi} \phi^*)~.}
The action \rf{g5} is the electromagnetic analog of \rf{d9}.   Note, however, that there is no coupling in \rf{g5} that grows in the extremal limit.   The near-horizon enhancement occurs here by a slightly different mechanism. In particular, due to the $ {1\over \sinh^2 z}$  factor the $dz$ integral diverges at the AdS$_2$ boundary $z\rt 0$. We therefore need to impose a cutoff $z=z_c$, meant to lie at the location where the AdS$_2$ approximation breaks down, and the geometry starts to go over into the asymptotically flat behavior. The precise location of the cutoff is not specified, though in terms of the radial coordinate $r$ it is clear that we should take $r_c \sim r_0$, which in terms of $z$ gives
\eq{g7}{ {1\over z_c} = {r_c -r_0 \over 2\pi r_0^2 T_H} \sim  {1\over 2\pi r_0 T_H}~.}
We then take 
\eq{g8}{ \int_{z_c}^\infty {dz \over \sinh^2 z} \approx {1\over z_c} \sim   {1\over 2\pi r_0 T_H}~.}
Using this, the  part of the action that contributes to the near-horizon enhancement is 
\eq{g9}{ S_{\rm en} = -\int\! dt \int_0^\infty dz  |\p \phi|^2 - {1\over z_c} \int\! dt  \int_0^\infty \! dz' J(z')\int_0^\infty \! dz'' J(z'') ~.}
We then see that the effective coupling is 
\eq{g10}{ g= {q^2\over z_c}  =   {q^2 \over 2\pi r_0 T_H}~,}
yielding the low-temperature enhancement. 

We can use this action to compute the tree-level four-point function. As in the gravitational case, to reduce the number of diagrams, we introduce two charged fields $V$ and $W$ and consider $\langle V^\dagger VW^\dagger W\rangle$. The corresponding action is 
\eq{g11}{ S_{\rm en} = -\int\! dt \int_0^\infty dz  \left(  |\p V|^2+|\p W|^2\right)  - {2\over z_c} \int\! dt  \int_0^\infty \! dz' J_V(z') \int_0^\infty \! dz'' J_W(z'') ~.}
We now compute the Euclidean signature on-shell action corresponding to the tree-level four-point function by contracting the quartic vertex with the massless bulk-boundary propagators.   The computation is similar to that in the gravitational case and results in
\eq{g12}{ S=  {2g \over (2\pi)^4} \sum_{k\neq 0}  \left[\sum'_{n_1+n_2=k} (|n_1|-|n_2|) e^{in_1\tau_1+in_2\tau_2}   \right] {1\over k^2} \left[\sum'_{n_3+n_4=-k} (|n_3|-|n_4|) e^{in_3\tau_3+in_4\tau_4}   \right]\,, }
where the $'$ on the summations instructs us to sum over only same-sign $(n_1,n_2)$ and same sign $(n_3,n_4)$. The factor of ${1\over k^2}$ indicates the existence of an exchanged field with propagator ${1\over k^2}$. We thus introduce  the phase variables $\theta(\tau) $, which is $2\pi$ periodic and has action 
\eq{g13}{ S_{\rm phase} = {1\over 2g} \int\! d\tau \dot{\theta}^2\,, }
which is the analog of the Schwarzian action.    We then couple this to dressed bilocal operators,
\eq{g14}{  S_{\rm bilocal}[\tau_i,\tau_j]  =   {e^{i(\theta(\tau_i) -\theta(\tau_j))} \over 2\pi \sin^2{\tau_{ij}\over 2}} ~.}
Then, if we compute
\eq{g15}{  \int\! [D\theta] e^{-S_{\rm phase} -S_{\rm bilocal}[\tau_1,\tau_2] -S_{\rm bilocal}[\tau_3,\tau_4] }  } 
to first order in $g$ by expanding each bilocal to first order in $\theta$ and then Wick contracting, we reproduce \rf{g12}. Using this  and the position space Wick contraction 
\eq{g15a}{ \langle \theta(\tau_i)\theta(\tau_j)  \rangle =  g  \left( {\pi \over 6}- {\tau_{ij}\over 2} +{\tau_{ij}^2 \over 4\pi} \right)~,\quad 2\pi > \tau_i > \tau_j> 0\,, }
we readily compute the four-point function with $ 2\pi >\tau_1>\tau_2 >\tau_3>\tau_4>0$ 
\eq{g16}{  \langle V^\dagger (\tau_1) V(\tau_2) W^\dagger(\tau_3) W(\tau_4)\rangle_{\rm en} & =- {1 \over (2\pi)^2} {g\over 2\pi } {\tau_{12} \tau_{34} \over \sin^2\left({ \tau_{12}\over 2}\right)  \sin^2\left({ \tau_{34}\over 2}\right)  }~.   }
As indicated, this is just the part of the four-point function proportional to the enhanced coupling $g$.

\section{\texorpdfstring{AdS$_3$}{AdS3} example}
\label{ads3}

We briefly discuss the case of AdS$_3$ gravity coupled to a real scalar field, reduced to the s-wave in the Hamiltonian formulation.  This case is technically simpler than the corresponding $d=3+1$ \RN example because, as is well known,  the s-wave reduction of AdS$_3$ gravity yields precisely JT gravity \cite{Achucarro:1993fd}.  The near-horizon limit in this case just corresponds to replacing the coupling of the scalar to the dilaton (size of the $S^1$) by its near-horizon form. 

The starting action is 
\eq{g17}{ S& =-  {1\over 2\pi} \int\! d^3x \sqrt{-g} \Big[ {1\over 8G_3}{\cal R} -{1\over 4G_3 \ell^2}  +{1\over 2} (\p \phi)^2 +V(\phi) \Big] }
and the s-wave ansatz is 
 \eq{g18}{ ds_3^2 = -N^2 dt^2 + L^2 (dr +N^r dt)^2 + R^2 d\psi^2~,\quad \phi = \phi(r,t)~. }
 As before, we choose the gauge $R=r$, $\pi_L=0$, and solve the constraints.    This leads to the following reduced scalar field action on the BTZ background
\eq{g19}{ S  = \int\! dt \int_{r_s}^\infty\! dr  \left( \pi_\phi \dot{\phi} -\Big(f(r)h(r) + rV(r)\Big)  e^{-\int_{r}^\infty \! dr' 8G_3h(r')}   \right) }
with 
\eq{g20}{ f(r) = {r^2 - r_h^2 \over \ell^2}~,\quad  h = {1\over 2} \left( {\pi_\phi^2 \over r}+r\phi'{}^2\right)~.}
The near horizon limit simply corresponds to making the replacement $ h \rt {1\over 2} \left( {\pi_\phi^2 \over r_h}+r_h\phi'{}^2\right)$.  Redefining variables as 
\eq{g21}{ r=r_h \coth z~,\quad t= {\ell^2 \over r_h} \tilde{t}~,\quad   \phi = {\phit\over \sqrt{r_h}}~,\quad  \pi_\phi = {\sinh^2 z \over \sqrt{r_h}} \tilde{\pi}_\phi~,}
and setting $V=0$, we arrive at the action 
\eq{g22}{ S & = \int\! d\tilde{t} \int_{0}^{z_s}\! dz \Big(  \tilde{\pi}_\phi \dot{\tilde{\phi}}  -  {1\over 2}  (\tilde{\pi}_\phi^2 + \tilde{\phi}'{}^2)  e^{-{g\over 2} \int_0^z\! dz' \sinh^2 z' (\tilde{\pi}_\phi^2 + \tilde{\phi}'{}^2) )}   \Big)  }
which is of the same form as \rf{c3}  (with $\tilde{m}=0$) but now with  the effective coupling 
\eq{g23}{ g = {2G_3 \over r_h} = { G_3 \over \pi \ell^2 T_H}~.  }
We again see the emergence of an effective coupling that grows as $T_H\rt 0$.  However, this is less physically relevant (even apart from the spacetime dimensionality) than for the \RN case.   At low temperatures, the horizon radius $r_h\rt 0$, which invalidates the s-wave reduction, unlike for the \RN case where the horizon size stabilizes at the extremal horizon radius $r_0$. We also note that in the canonical ensemble at low temperatures the BTZ black hole is thermodynamically disfavored compared to thermal AdS$_3$; however, this does not preclude consideration of a low-mass black hole being formed from the collapse of a shell; it just says that such a black hole will radiate and eventually decay into thermal AdS$_3$. In any case, if we (without justification) use the s-wave reduction, then we will obtain the same AdS$_2$  boundary correlators as before since the action is the same.

\bibliographystyle{bibstyle2017}
\bibliography{collection}

\end{document}